\documentclass[%
reprint,
superscriptaddress,
amsmath,amssymb,
aps,
]{revtex4-2}

\usepackage{graphicx}
\usepackage{dcolumn}
\usepackage{bm}
\usepackage{soul}
\usepackage{float}
\usepackage{tabularx}
\usepackage{array}
\usepackage{multirow}
\newcolumntype{Y}{>{\centering\arraybackslash}X}
\usepackage{siunitx}
\usepackage{hyperref}
\hypersetup{colorlinks=true, citecolor=blue, urlcolor=blue, linkcolor=blue}
\usepackage{todonotes}
\usepackage{xcolor}
\usepackage{colortbl}
\makeatletter
\usepackage{braket}
\def\CT@@do@color{%
	\global\let\CT@do@color\relax
	\@tempdima\wd\z@
	\advance\@tempdima\@tempdimb
	\advance\@tempdima\@tempdimc
	\advance\@tempdimb\tabcolsep
	\advance\@tempdimc\tabcolsep
	\advance\@tempdima2\tabcolsep
	\kern-\@tempdimb
	\leaders\vrule
	\hskip\@tempdima\@plus  1fill
	\kern-\@tempdimc
	\hskip-\wd\z@ \@plus -1fill }
\makeatother
\usepackage{colortbl}

\usepackage{soul}
\usepackage{upgreek}

\begin{document}

\title[Three-Wave Mixing Quantum-Limited Kinetic Inductance\\ Parametric Amplifier operating at 6~Tesla and near 1~Kelvin]%
{Three-Wave Mixing Quantum-Limited Kinetic Inductance\\ Parametric Amplifier operating at 6~Tesla and near 1~Kelvin}

\author{S.~Frasca}
\email[E-mail: ]{simone.frasca@epfl.ch}
\affiliation{Institute of Physics, Swiss Federal Institute of Technology (EPFL), 1015 Lausanne, Switzerland.}		
\affiliation{Center for Quantum Science and Engineering, EPFL, 1015 Lausanne, Switzerland.}

\author{C.~Roy}
\affiliation{Institute of Physics, Swiss Federal Institute of Technology (EPFL), 1015 Lausanne, Switzerland.}		
\affiliation{Center for Quantum Science and Engineering, EPFL, 1015 Lausanne, Switzerland.}

\author{G.~Beaulieu}
\affiliation{Institute of Physics, Swiss Federal Institute of Technology (EPFL), 1015 Lausanne, Switzerland.}		
\affiliation{Center for Quantum Science and Engineering, EPFL, 1015 Lausanne, Switzerland.}

\author{P.~Scarlino}
\affiliation{Institute of Physics, Swiss Federal Institute of Technology (EPFL), 1015 Lausanne, Switzerland.}		
\affiliation{Center for Quantum Science and Engineering, EPFL, 1015 Lausanne, Switzerland.}

\date{\today}

\begin{abstract}
Parametric amplifiers play a crucial role in modern quantum technology by enabling the enhancement of weak signals with minimal added noise. Traditionally, Josephson junctions have been the primary choice for constructing parametric amplifiers. Nevertheless, high-kinetic inductance thin films have emerged as viable alternatives to engineer the necessary nonlinearity.
In this work, we introduce and characterize a Kinetic Inductance Parametric Amplifier (KIPA) built using high-quality NbN superconducting thin films. 
The KIPA addresses some of the limitations of traditional Josephson-based parametric amplifiers, excelling in dynamic range, operational temperature, and magnetic field resilience. 
We demonstrate a quantum-limited amplification ($>20$~dB) with a 20~MHz gain-bandwidth product, operational at fields up to 6~Tesla and temperatures as high as 850~mK. 
Harnessing kinetic inductance in NbN thin films, the KIPA emerges as a robust solution for quantum signal amplification, enhancing research possibilities in quantum information processing and low-temperature quantum experiments. Its magnetic field compatibility and quantum-limited performance at high temperatures make it an invaluable tool, promising new advancements in quantum research.

\end{abstract}

\maketitle

\subsection*{\label{sec:intro}Introduction}


Parametric amplifiers (PAs) are essential tools for enhancing and consequently detecting weak quantum signals, including single photons and complex quantum states, which are central to the exploration of quantum phenomena \cite{Caves_1982, Aumentado_2020}.
These signals contain crucial information about the studied quantum systems.  
However, their inherently low intensity poses a significant challenge for accurate measurement \cite{Clerk_2010}. 
Within the microwave frequency spectrum, parametric amplifiers facilitate precise, rapid, and high-fidelity single-shot measurements for a broad spectrum of systems, including superconducting qubits \cite{walter_2017, Vijay_2011}, spin qubits based on quantum dots \cite{Stehlik_2015, Schaal_2020, Elhomsy_2023}, spin ensembles \cite{bienfait_2016, eichler_2017, wang_2023, Vine_2023}, and nanomechanical resonators \cite{Kerckhoff_2013, Clark_2016}.

While cryogenic amplifiers based on CMOS or HEMT technology thermally anchored at 4 K have traditionally been employed in most microwave readout architectures in quantum sensing and computing
\cite{peng_2021}, they introduce substantial noise, which in turn impacts the fidelity of single-shot readout processes \cite{walter_2017}. 
In stark contrast, parametric photon conversion offers a highly effective alternative for achieving substantial amplification with minimal added noise, ultimately constrained by the fundamental principles of quantum mechanics \cite{Clerk_2010}.

Beyond their primary function of amplification, parametric amplifiers emerge as invaluable tools for generating non-classical radiation states. Specifically, they excel in producing squeezed coherent radiation or squeezed noise below the vacuum level \cite{Movshovich_1990, Castellanos-Beltran_2008, Malnou_2019}. 
Quantum squeezed states are of significant importance as they play a crucial role in increasing the signal-to-noise ratio of measurements. This improvement goes beyond the constraints of the standard quantum limit \cite{Giovannetti_2004}, enabling enhanced detection sensitivity \cite{Malnou_2019} and facilitating the implementation of secure quantum communication protocols \cite{pogorzalek_2019}.


At the core of parametric amplification lies the exchange of energy between two distinct tones: the pump and the signal, taking place within a nonlinear medium \cite{Stolen_1982}. 
To efficiently convert energy from the pump to the signal and achieve a significant parametric gain, it is essential to maximize the interaction time with the nonlinear medium.
Parametric amplifiers can be categorized into two primary types based on the approach used to extend this interaction time: resonant-type and traveling wave-type amplifiers \cite{Aumentado_2020}. 
Resonant amplifiers achieve prolonged interaction times by using resonant structures, while traveling wave-type amplifiers (TWPAs \cite{Macklin_2015}) allow the co-propagation of the pump and signal through a long, nonlinear transmission line. 
In general, TWPAs offer broader bandwidths compared to resonant amplifiers \cite{Esposito_2021}. However, it is worth noting that TWPAs tend to introduce slightly more noise, often in the range of single-digit photon numbers, compared to their resonant counterparts \cite{Aumentado_2020}, which typically operate closer to the quantum limit with an addition of approximately 0.5 photons \cite{Castellanos-Beltran_2007}.


Both categories of parametric amplifiers can operate in either the four-wave mixing (4WM) or three-wave mixing (3WM) modes, determined by the device's order of nonlinearity. In 4WM mode, the amplification process involves four photons ($2\hbar\omega_p = \hbar\omega_s + \hbar\omega_i$), whereas 3WM mode amplification relies on a three-photon interaction ($\hbar\omega_p = \hbar\omega_s + \hbar\omega_i$), where $\omega_k$ for $k = p, s, i$ denotes the frequencies of the pump, signal, and idler waves, respectively \cite{Malnou_2021}. 
The three-wave mixing operation offers enhanced efficiency in separating the pump and signal tones, simplifying the pump filtering process\cite{yamamoto_2008}.


The most widely adopted microwave parametric amplifiers leverage the inherent nonlinearity of Josephson junctions. Josephson junction-based parametric amplifiers (JPAs \cite{Castellanos-Beltran_2007} and JTWPAs \cite{Macklin_2015}) have demonstrated exceptional quantum-limited or near-quantum-limited parametric amplification capabilities. 
However, these amplifiers necessitate magnetic shielding due to their extreme sensitivity to even minor magnetic fields, primarily arising from the low critical field of aluminum Josephson junctions \cite{Kuzmin_2023}.
The presence of a magnetic field significantly degrades the performance of Josephson junctions, imposing strong constraints on their compatibility with other quantum technologies, such as spin qubits \cite{loss_divincenzo_1998} and spin ensemble quantum memories \cite{bienfait_2016,eichler_2017}, which heavily rely on external magnetic fields for operation. This inherent susceptibility poses substantial challenges in realizing on-chip integration of Josephson junction amplifiers \cite{Elhomsy_2023}, which would enable the reduction of cable losses.


Recent advancements have brought forth significant improvements in parametric amplification techniques that exploit the nonlinearity inherent in kinetic inductance (high-$L_k$) in thin "dirty" superconducting films \cite{annunziata_tunable_2010}. Both resonant \cite{Tholen_2007} and traveling-wave \cite{day_paramp_2012} configurations have been successfully demonstrated. Notably, a significant breakthrough has been reported, demonstrating the use of $dc$ bias current to activate a three-wave mixing process in Kinetic Inductance Parametric Amplifiers (3WM-KIPA) \cite{Malnou_2021, parker_2022, Vine_2023}. However, until very recently, relatively little attention has been directed towards studying the magnetic field resilience of these systems \cite{Xu_2023, Khalifa_2023}.

\begin{figure*}	
    \centering
    \includegraphics[width=.96\linewidth]{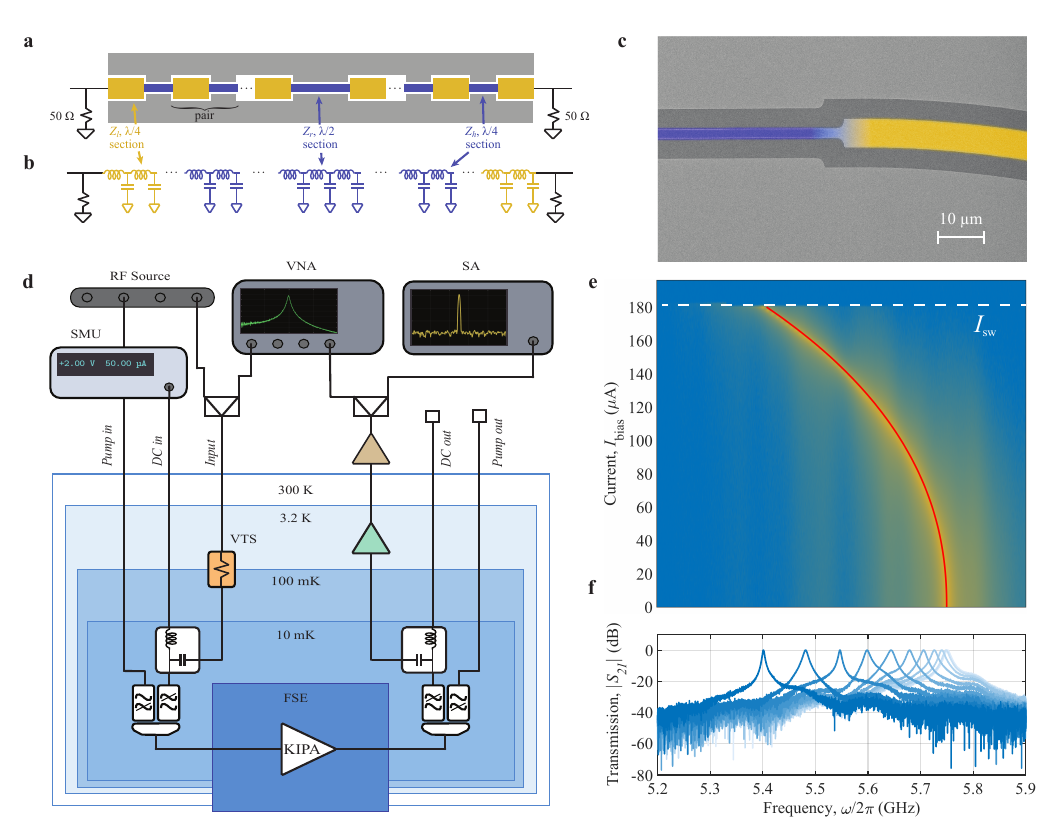}
	\caption{\textbf{Device and Experimental setup.} \textbf{(a-b) Device schematic}. The amplifier is integrated into a stepped impedance filter (SIF) and coupled to 50~$\Omega$ microwave ports in a Fabry-P\'erot configuration. The SIF consists of cascaded $\lambda/4$ CPW sections of impedance $Z_l$ (yellow) and $Z_h$ (blue) of 450 $\Omega$ and 900 $\Omega$ respectively. \textbf{(c) Device SEM}. False-color scanning electron micrograph of the device sections $Z_h$ (in blue) and $Z_l$ (in yellow). \textbf{(d) Simplified experimental setup}. Schematics representation of the experimental setup used to characterize the device in transmission. Signal, pump and $dc$ bias current are combined at base temperature using a bias tee and a diplexer [see Appendix~\ref{app:setup} for further details]. The Variable Temperature Source (VTS) is used for characterizing the KIPA noise temperature. SMU, VNA and SA stand for source/measurement unit, vector network analyzer and spectrum analyzer respectively. \textbf{(e-f) DC characterization}. Magnitude response $|S_{21}|$ of the device in transmission at various bias currents $I_\text{bias}$. We measure the resonant frequency shift of the device as function of $I_\text{bias}$, and extracted a switching current $I_\text{sw}$ of 182~$\upmu$A [dashed white line] and a critical current $I_*$ of 345~$\upmu$A [red line, representing the fit of Eq.~\eqref{eq:clem}].}
	\label{fig:setup}
\end{figure*}

The nonlinearity responsible for parametric amplification in high-$L_k$ thin films arises from the relation between kinetic inductance and current \cite{clem_geometry_2011}. This expression is typically approximated as a quadratic term of nonlinearity, denoted as $L_{k,0} [ 1 + ( I/I_* )^2]$, where $L_{k,0}$ represents the 
zero-current kinetic inductance and $I_*$ is the critical current. 
This term is analogous to the nonlinearity observed in optical Kerr media, known as self-Kerr, which is responsible for four-wave mixing processes. However this relation, in the presence of both direct and high-frequency current components, modifies into:
\begin{equation}
L_k = L_{k,0} \bigg[ 1 + \frac{I_{dc}^2}{I_*^2}  + \frac{2 I_{dc} I_{\mu w}}{I_*^2} + \frac{I_{\mu w}^2}{I_*^2} \bigg]. \label{eq:LkI}
\end{equation}
Here, $I_{dc}$ and $I_{\mu w}$ represent the contributions of direct current and microwave current.

The classical Kerr Hamiltonian, which can be described as
\begin{equation}
H_{\text{Kerr}}/\hbar = \omega_r a^\dagger a + K a^\dagger a + \frac{K}{2} a^\dagger a^\dagger a a, \label{eq:Hkerr}
\end{equation}
where $\omega_r$ is the resonant frequency and $K$ is the self-Kerr term, can be modified by the introduction of a $dc$ current $I_{dc}$ to become the Hamiltonian for the kinetic inductance parametric amplifier (or KIPA):
\begin{equation}
H_{\text{KIPA}}/\hbar = \Delta a^\dagger a + \frac{\xi}{2} a^{\dagger 2} + \frac{\xi^*}{2} a^2 + \frac{K}{2} a^\dag a^\dag a a. \label{eq:Hkipa}
\end{equation}
Here, $\Delta = \omega_r + \delta_{dc} + \delta_{\mu w} + K - \omega_p/2$, $\xi$ is the three-wave mixing pump strength defined as \cite{parker_2022}
\begin{equation}
    \xi = - \frac{1}{4} \frac{I_{dc}~ I_{\mu w}}{I_*^2} \omega_r e^{-i\psi_p}, \label{eq:3wm}
\end{equation}
where $\psi_p$ is the phase difference between the pump and signal tones.
The presence of a $dc$ current component in Eq.~\eqref{eq:Hkipa}
effectively reduces the degree of nonlinearity of the system, introducing new terms that are linear in the number of photons in the Hamiltonian, thus leading to three-wave parametric processes \cite{Malnou_2021}.

Compared to the Josephson parametric amplifiers, KIPAs offers several advantages. In particular, some thin films disordered superconductors such as NbN and NbTiN have already proven exceptional
resilience to in-plane magnetic field \cite{yu_2021, samkharadze_2016}. Furthermore, they also present an high critical temperature $T_C$ \cite{frasca_2023}, low fabrication complexity, and a significantly lower nonlinearity compared to Josephson junctions \cite{Khalifa_2023}. A reduced nonlinearity results in a larger critical current, 
expanding the
dynamic range and saturation power \cite{parker_2022}, all at the expenses of a typically larger static power dissipation. 
In this context, operating a KIPA with 3WM processes simplifies pump filtering, preventing potential
saturation or even damage to subsequent amplifiers in the amplification chain.


In this study, we introduce a quantum-noise-limited 3WM resonant KIPA based on high-$L_k$ NbN thin films and assess its performance under various conditions, including large in-plane magnetic fields and high temperatures. 
Following an overview of the device and its performance characterization under conventional operational conditions (low temperature and zero field), we investigate how the KIPA's key properties, such as noise characteristics, gain-bandwidth product, and saturation power, evolve with changes in magnetic field and temperature. 
The amplifier presents a saturation power of $-66$~dBm, a gain-bandwidth product of approximately 20~MHz, and quantum-noise limited noise performance across diverse operational conditions. These exceptional NbN KIPA performance characteristics are maintained even in the presence of in-plane magnetic fields up to 6~Tesla (the maximum field available in our experiment) and operating temperatures of up to 850~mK, slightly above 15\% of $T_C$.

\subsection*{\label{sec:design}Device design}

The device is structured as a half-wave, coplanar waveguide resonator embedded into a stepped-impedance filter (SIF, \cite{bronn_2015, Sigillito_dbr_2017, liu_dbr_2017}), as illustrated in Fig.~\ref{fig:setup}(a-c). Our design is fully compatible with readout configurations in either transmission or reflection modes. This additional layer of flexibility can prove extremely beneficial for chip integration of parametric amplifier operating in high magnetic fields, for instance when coupled to spin qubits. 
This is particularly advantageous, as conventional resonant parametric amplifiers used in reflection necessitate microwave circulators in the amplification chain, many of which are not compatible with magnetic fields.

In the attempt of reducing the static power dissipation of the KIPA, we utilize NbN as the nonlinear medium due to its lower critical current density and higher sheet kinetic inductance ($L_{k,\square}$) compared to the more commonly used NbTiN films \cite{parker_2022, Malnou_2021}. These characteristics result in enhanced self-Kerr nonlinearity and enable a more compact device design. 
Our films are deposited via bias-sputtering \cite{dane_bias_2017, frasca_2023} on high-resistivity Si wafers ($\rho \ge 10~k\Omega$cm) with a thickness of 13~nm, resulting in a sheet kinetic inductance of $L_{k,\square} = 100$~pH/$\square$ and typical critical temperature $T_C = 5.6$~K. The devices are patterned with e-beam lithography and etched with fluorine chemistry in a reactive ion etcher. Further details about device fabrication can be found in Appendix~\ref{app:materials}. 

As detailed in Refs.~\cite{parker_2022, Vine_2023}, the inclusion of the stepped-impedance filter (SIF) allows to partially decouple the resonator from the feedline at high-frequency while maintaining a galvanic connection, needed to apply the necessary $dc$ bias to the resonator for establishing 3WM processes.
By embedding the resonator between two SIFs, we effectively create a Fabry-P\'erot cavity, with the SIFs acting as the microwave equivalent of Bragg mirrors \cite{pozar_2012}. 
The SIFs are composed of a series of quarter-wave sections alternating between high and low impedance segments [see Fig.\ref{fig:setup}(a-b)], designed to create a stopband at the central frequency of the KIPA. The external coupling coefficient $\kappa$ for this Fabry-P\'erot cavity is determined as follows [refer to Appendix~\ref{app:coupling}]:
\begin{equation}
    \kappa = \frac{\pi}{4} \frac{Z_{\text{eff}}}{Z_r} \omega_0 = \frac{\pi}{4} \frac{Z_l^{2 n_l}}{Z_h^{2 n_h}} \frac{1}{Z_r Z_0} \omega_0, \label{eq:kSIF}
\end{equation}
where $\omega_0$ represents the common resonant frequency of all sections of the SIF and the main cavity, $Z_r$ corresponds to the impedance of the resonator, $Z_0$ is the environment impedance, $Z_l$ [$Z_h$] is the low [high] impedance segment of the SIF, and $n_l$ [$n_h$] is the number of low [high] impedance sections within the SIF.

\begin{figure*}
    \centering
    \includegraphics[width=.96\linewidth]{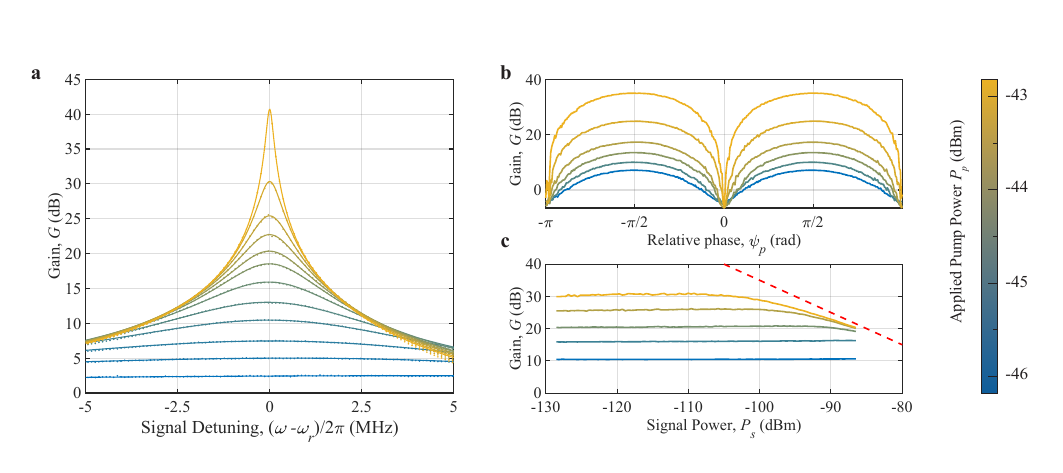}
	\caption{\textbf{KIPA characterization at $T_\text{dev}=10~$mK and $B_\parallel=0$~T}. \textbf{(a) Gain profile and gain-bandwidth product} Phase-insensitive gain of the KIPA as function of signal frequency $\omega$ for different pump powers (as indicated by the colorscale). The data are fitted with Eq.~\eqref{eq:gain} to extract the gain-bandwidth product. \textbf{(b) Degenerate operation}. Phase-sensitive gain behavior obtained operating at $\omega_s = \omega_p / 2$ and modulating the pump phase $\psi_p$. The phase response has been aligned such that  $\psi_p = \pi/2$ corresponds to maximum gain. \textbf{(c) Saturation power}. Phase-insensitive saturation power measured as signal power dependent gain for different pump powers. The red dashed lines represents the $P_\text{1dB} = -65$~dBm saturation threshold.}
	\label{fig:characteristics}
\end{figure*}

To significantly enhance the overall device's resilience to both parallel and perpendicular magnetic fields, we design the stepped impedance filter using coplanar waveguide sections with thin and narrow center conductor.
Consequently, we achieve relatively large impedance of 450~$\Omega$ and 900~$\Omega$ for the $Z_l$ and $Z_h$ sections, respectively. Given the large number of SIF pairs (5.5 on each side of the cavity), and considering the cavity's impedance as $Z_r = 900~\Omega$ and its resonant frequency $\omega_0/2\pi = 5.75$~GHz, we expect an external coupling rate $\kappa/2\pi$ of 19.8~MHz, corresponding to a coupling quality factor $Q_c \approx 290$. 

\subsection*{\label{sec:characterization}Characterization at optimal conditions}

The measurements are conducted inside a BlueFors LD250 dilution refrigerator equipped with a fast sample exchange system (FSE), as illustrated in Figure ~\ref{fig:setup}(d). A first estimation of the critical current of the device is obtained by monitoring the frequency shift of the transmission peak of the KIPA with respect to $dc$ bias [see Fig.~\ref{fig:setup}(e-f)]. The central resonator functions as a half-wave resonator, exhibiting peak transmission at its resonant frequency. By tracking the frequency shift of the transmission peak and considering the relation \cite{clem_kinetic_2012, frasca_2019}:
\begin{equation}
\frac{L_k(I)}{L_k(0)} = \bigg[ 1 - \bigg( \frac{I}{I_*} \bigg)^n \bigg]^{-1/n}, \label{eq:clem}
\end{equation}
with $n = 2.21~\text{for}~T/T_C < 0.1$, we determine a resonator critical current $I_* = 345~\upmu$A [see red line in Fig.~\ref{fig:setup}(e)]. At the same time, the maximum operational bias current, also known as switching current ($I_\text{sw}$), is as large as 180~$\upmu$A. This large ratio of switching-to-critical current highlights the high quality and uniformity of the deposited superconducting film \cite{frasca_2019}, enabling the tuning of the KIPA frequency up to 400~MHz, equivalent to approximately 7\% of the KIPA resonant frequency.

To minimize the contributions of higher-order nonlinearities, we characterize the amplifier at a static bias condition of $I_{b} = 80~\upmu$A. Lower $I_{b}$ would require higher pump power to achieve the desired amplification levels. We estimate the four-wave mixing strength (\textit{i.e.} the self-Kerr nonlinearity, $K$) as:
\begin{equation}
    K = -\frac{3}{8} \frac{\hbar \omega_r^2}{L_t I_*^2} = -0.133~\text{Hz}, \label{eq:4wm}
\end{equation}
given the total resonator inductance $L_t = 82.4$~nH. This value is significantly smaller than the cavity coupling $\kappa$, satisfying the condition $\kappa/ |K| \ll 1$. Therefore, as $|\xi| \to \kappa$, which is the condition that maximizes 3WM gain [see Appendix~\ref{app:iotheory}], the KIPA parametric process is predominantly dominated by 3WM \cite{Malnou_2021}.

To determine the optimal pump frequency, we monitor the KIPA gain in the vicinity of half pump frequency using a relatively low pump power to prevent the induction of self-oscillations (which can occur when $|\xi| > \kappa$ \cite{Vine_2023b}). 
We set the offset between the probe frequency and the half pump frequency to guarantee non-degenerate amplifier operation, ensuring insensitivity to pump phase. 
Interestingly, we observe that in our system, the optimal pump frequency is slightly redshifted compared to the transmission peak. 
This deviation arises from our design choice of using similar conductor widths for the SIF and the central resonator, making the device more resilient to magnetic fields, a key objective of our study.
However, this design choice also leads to a situation where the application of a $dc$ current to power the device results in higher current density in the $Z_h$ sections compared to the $Z_l$ sections due to their narrower conductor widths. Consequently, the resonant frequency of the $Z_h$ sections shifts more rapidly than that of the $Z_l$ sections. As a result, the resonant frequency of the main cavity ends up being offset from the transmission peak of the Fabry-P\'erot cavity, resulting in insertion loss. While this effect is suboptimal for amplifier operation, a straightforward solution would be to design the central resonator and the high-impedance sections at a slightly higher frequency than the low-impedance sections. This adjustment would compensate for the frequency mismatch caused by $dc$ biasing.


When biasing the KIPA with a current of $I_{b} = 80~\upmu$A, the transmission peak shifts to $\omega_\text{peak}/2\pi = 5.689$~GHz, while the optimal pump condition is identified at $\omega_p/2\pi = 11.347$~GHz, implying a central resonator frequency of $\omega_r/2\pi = 5.6735$GHz. This difference results in an insertion loss in transmission of approximately 7~dB, equivalent to $10 \log(\eta_{IL})$ of $-3.5$~dB on each side of the main resonator. This insertion loss will significantly impact the noise temperature estimation of the device, as will be demonstrated later.

A detailed explanation of the measurements setup is provided in Appendix~\ref{app:setup}. The gain characteristics of the KIPA are presented in Fig.~\ref{fig:characteristics}(a) and  studied across increasing pump powers, ranging from $-46$ to $-43$~dBm at the device input. The data are fitted to the input-output relation \cite{parker_2022} adapted for cavity measured in transmission, leading to the signal gain equation [see Appendix~\ref{app:iotheory}]:
\begin{equation}
    g(\omega) = \frac{\kappa (2 \kappa + \gamma)/2 - i \kappa (\Delta + \omega - \omega_p/2)}{\Delta^2 + [(2 \kappa + \gamma)/2 - i(\omega - \omega_p/2)]^2 - |\xi|^2}. \label{eq:gain}
\end{equation}
The gain-bandwidth product, $GBP~=~\sqrt{G}~\delta\omega_{\text{3dB}}~\approx~\kappa$, where $G$ is the power gain and $\delta\omega_{\text{3dB}}$ is the 3~dB bandwidth of the amplifier, is estimated to be $21.3 \pm 0.6$~MHz, in excellent agreement with the expected value [see Appendix~\ref{app:coupling}]. 


KIPAs can also be operated as phase-sensitive amplifiers in the so-called degenerate mode \cite{parker_2022}. When the signal corresponds to exactly half the pump frequency, \textit{i.e.} when $\omega_s = \omega_p/2$, the signal and idler tones interact, either constructively or destructively. This phenomenon gives rise to the capability of the device as a phase-sensitive amplifier, referred to as degenerate amplification. In this condition, following the same formalism presented in Ref.~\cite{parker_2022}, Eq.~\eqref{eq:gain} becomes 
\begin{equation}
    |g(\psi_p)| = \bigg | \frac{\kappa (2 \kappa + \gamma)/2 + i \kappa \Delta + i \kappa |\xi| e^{-i\psi_p}}{\Delta^2 + [(2 \kappa + \gamma)/2 ]^2 - |\xi|^2} \bigg |, \label{eq:degenerate}
\end{equation}
where $\psi_p$ is the relative phase between the pump and signal tones. The phase-amplification characteristic of our KIPA is reported in Fig.~\ref{fig:characteristics}(b). Operating at a relative phase $\psi_p = \pi/2$ [$\psi_p = 0$], more than 30~dB of gain [nearly 6~dB of deamplification] is achieved.

Another important figure of merit of parametric amplifiers is their power handling capabilities, also known as saturation, which is characterized by the so-called one~decibel (1-dB) compression point. We measure the 1-dB compression point by modulating the signal power in non-degenerate mode [see Fig.~\ref{fig:characteristics}(c)], \textit{i.e.} for $\omega_s = \omega_p/2 + \delta\omega$, with $\delta\omega = 10$~kHz small enough to ensure non-degenerate operation. At a gain level of roughly 21~dB, we observe 1-dB compression point of approximately $-86$~dBm, leading to a saturation power of roughly $P_\text{1dB} = -65$~dBm. This saturation power is roughly three orders of magnitude larger than the saturation power of typical Josephson junction parametric amplifiers \cite{planat_2019}. We expect to increase this value up to 6~dB by operation the device in degenerate mode at $\psi_p = \pi/2$.

\begin{figure*}
    \centering
    \includegraphics[width=.96\linewidth]{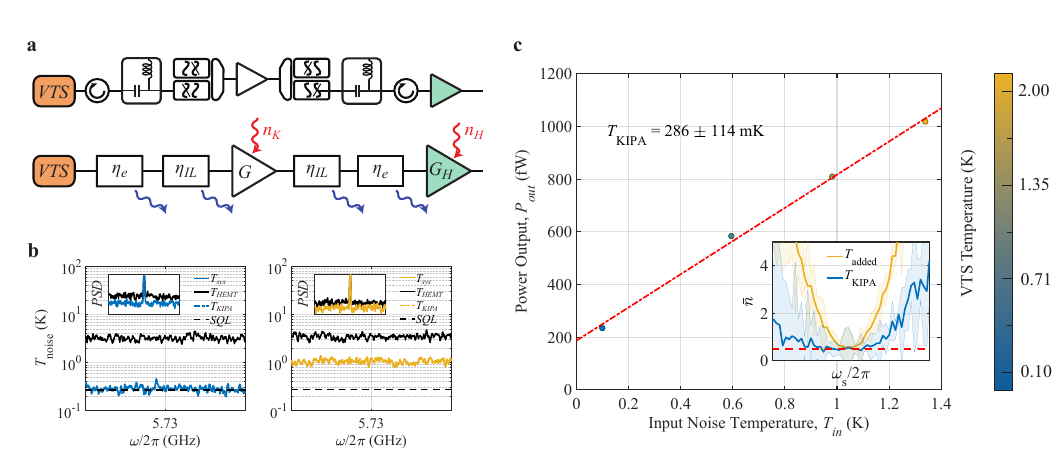}
	\caption{\textbf{Noise.} \textbf{(a) Noise measurement schematics}. Top: schematic representing the cryogenic components in series between the VTS and the HEMT [see Appendix~\ref{app:setup} for further details]. Bottom: equivalent diagram of the cascaded transmission efficiencies
    and gains. $\eta_e$ [$\eta_{IL}$] is the global attenuation of the cryogenic components [is the insertion loss of the KIPA]; $G$ [$G_H$] represents the KIPA [HEMT] gain. \textbf{(b) Noise floor acquisition}. All noise measurements are conducted using a spectrum analyzer operating with a bandwidth of 100~Hz, averaging the noise power spectral density over a range of 10~kHz. The power output is then converted in noise temperature according to Eq.~\eqref{eq:pout}. In the insets: measured power spectral density in presence of a coherent tone to calibrate the KIPA gain. In blue [yellow] the noise spectra acquired with VTS temperature of 100~mK [2~K], highlighting the difference in noise floor. \textbf{(c) KIPA noise temperature}. Power output [measured according to panel (b)] in correspondence of different VTS temperatures. The red dashed line represents a linear fit according to Eq.~\eqref{eq:pout} to estimate the added noise $T_\text{add}$. The extracted KIPA noise temperature at maximum gain frequency is $T_\text{KIPA} = 286 \pm 114$~mK,  which is remarkably close to the quantum limit of half a photon. Inset: estimated equivalent noise photon number as function of frequency. The shaded areas represent the error associated to the estimate, taking into account insertion loss asymmetry as explained in Appendix~\ref{app:tnoise}. The dashed red line represents the half photon quantum limit.} 
	\label{fig:noise}
\end{figure*}

\subsection*{\label{sec:noise}Noise performance}

An essential aspect concerning parametric amplifiers, as previously mentioned, relates to noise performance. Resonant parametric amplifiers based on Josephson junctions conventionally approach the quantum noise limit of 0.5~added photons when functioning in non-degenerate mode. In contrast, the achievement of such capabilities in KIPAs is relatively recent \cite{Xu_2023, Khalifa_2023}.

In evaluating our amplifier's noise characteristics, we perform noise thermometry calibration with a commercial variable temperature source (VTS), as indicated in Fig.~\ref{fig:setup}(d), supplied by BlueFors \cite{Simbierowicz_2021}. The VTS is mounted to the 100~mK stage of the cryostat, thermally weakly connected and far from the sample holder to prevent radiative and conductive heat exposure to the device. 
First, we perform an accurate calibration of the losses associated with the microwave components and cables placed between the VTS and the cryogenic HEMT low-noise amplifier [see Fig.~\ref{fig:noise}(a-b)]. Then, we sweep the temperature of the VTS from 100~mK to 2~K in four steps, which is used to fine control the input noise power radiated to the device $I_\text{in}$. Following the same protocol described in Ref.~\cite{Simbierowicz_2021}, we monitor the output power $P_\text{out}$ with a spectrum analyzer, operated at $B = 100$~Hz bandwidth and averaging 20 waveforms to reduce measurement noise. A detailed quantification of $T_\text{in}$, taking into account cryogenic components losses ($\eta_{e}$) and device insertion loss ($\eta_{IL}$), is reported in Appendix~\ref{app:tnoise}. 

The KIPA noise temperature $T_\text{KIPA}$ is extracted according to
\begin{equation}
    \frac{P_\text{out}}{G_{\text{tot}} k_B B} = T_\text{in} + T_\text{add} + \frac{T_{\text{bkg}}}{G_{\text{HEMT}}~G}, \label{eq:pout}
\end{equation}
where: $T_\text{add} = T_{\text{KIPA}} + T_{\text{HEMT}}/G$ is the added noise to the measurement extracted by fitting the output power relation; $T_\text{HEMT} = 1.95$~K is the calibrated HEMT noise temperature closely aligned to the value provided by the factory [see Appendix~\ref{app:tnoise}]; $T_\text{bkg}$ is the noise temperature of the room temperature electronics; $G_\text{HEMT}$ [$G$] is the HEMT [KIPA] gain at the specific frequency under study; $G_\text{tot}$ is the total gain from the output of the KIPA to the spectrum analyzer, calibrated to be 68.2~dB [see Appendix~\ref{app:setup}]; and $B$ is the integration bandwidth of the spectrum analyzer. 
Fig.~\ref{fig:noise}(b) provides insight into the increased noise level depending on the VTS temperature, as observed through measurements conducted with the spectrum analyzer. As shown from Fig.~\ref{fig:noise}(c), the extracted noise temperature of the KIPA is $286 \pm 114$~mK, equivalent to $1.04 \pm 0.41$ photons, of which 0.5 are from the vacuum and $0.54 \pm 0.41$ are added from the parametric amplifier, approaching the quantum limit. Similar noise performance have been extracted when operating the amplifier in reflection configuration [see Appendix~\ref{app:still}].


\begin{figure}
    \centering
    \includegraphics[width=\linewidth]{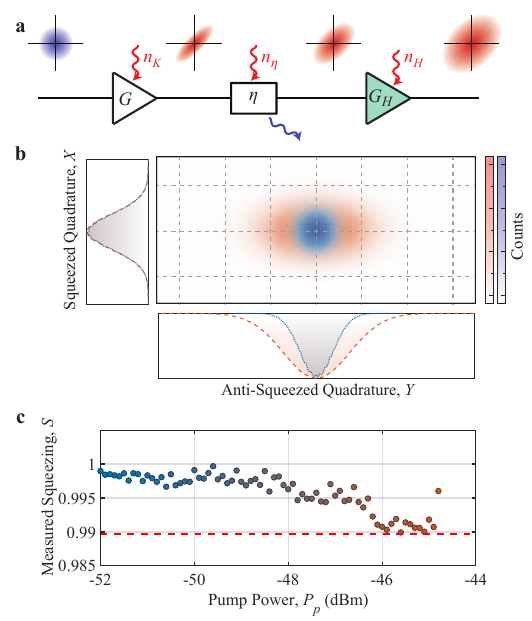}
	\caption{\textbf{Vacuum Squeezing}. \textbf{(a) Squeezing schematics}. Diagram of cascaded gain and attenuation and equivalent schematic depicting the evolution of a squeezed state of vacuum. $n_i$ represents the added photon number by the $i$-th element. \textbf{(b) Squeezed vacuum}. Measured vacuum (blue) and squeezed vacuum (red) states. Histogram of the $X$ and $Y$ quadratures signals measured at room temperature with the spectrum analyzer. Also in blue and in red, the distributions along the two quadratures for vacuum and squeezed states, respectively. \textbf{(c) Measured squeezing $S$}. Squeezing value $S$, as defined in Eq.~\eqref{eq:squeezing}, as function of pump power $P_{p}$. Taking into account both losses ($\eta$) and HEMT noise ($n_\text{H}$), we derive the maximum measurable squeezing $S|_\text{min} = 0.9897$ (red dashed line). From Eq.~\eqref{eq:squeezing}, we extract a maximum deamplification gain $G_X = -2.95$~dB. The last point at $P_p = -44.8$~dBm shows the triggering of the self-oscillations described in Ref.~\cite{Vine_2023b}.}
	\label{fig:squeeze}
\end{figure}

\subsection*{\label{sec:squeezing}Squeezing below vacuum}

A perfect degenerate parametric amplifier is a device capable, at its degenerate frequency $\omega = \omega_p/2$, to amplify one of the signal quadratures and deamplify the other without introducing noise \cite{Vaartjes_2023}. Given that our device is both quantum noise-limited and exhibits substantial phase-sensitive amplification [see Fig.~\ref{fig:characteristics}(b)], we investigate the feasibility of generating squeezed states below the vacuum. 

The amplification process of the vacuum state due to the KIPA is depicted in Fig.~\ref{fig:squeeze}(a). Any signal output from the KIPA is subject to attenuation $\eta = \eta_e \times \eta_{IL}$ before being further amplified by the HEMT low-noise amplifier, due to lossy cables and cryogenic components between the two amplifiers, as shown in Fig.~\ref{fig:noise}(a). To quantify the extent of squeezing in the parametric amplifier, we examine the two quadratures of the incoming vacuum field as
\begin{equation}
    X_\text{in} = \frac{\hat{a} + \hat{a}^\dagger}{2}, \quad Y_\text{in} = \frac{\hat{a} - \hat{a}^\dagger}{2i}.
\end{equation}

We conduct the preparation of a vacuum squeezed state by turning on the KIPA in degenerate mode and measuring the two quadratures of the vacuum at room temperature. In Fig.~\ref{fig:squeeze}(b), we present the distributions of the squeezed vacuum quadratures (depicted in red) and compare them to the distributions of the normal vacuum (depicted in blue), \textit{i.e.} the signal obtained with the amplifier turned off. The variance of the squeezed quadrature \cite{Vaartjes_2023} measured at room temperature is quantified as [see Appendix~\ref{app:iotheory}]
\begin{align}
    \braket{X^2_\text{out}} = & G_\text{H} G_X \eta \braket{X^2_\text{in}} + G_\text{H} (G_X-1) \eta n_{\kappa,2} + \nonumber \\ & G_\text{H} G_X \eta n_{\gamma} + G_\text{H}(1-\eta) \bigg( \frac{1}{4} + \frac{n_{\eta}}{2} \bigg) + \nonumber \\ & (G_\text{H}-1)n_\text{H}.
\end{align}
In this equation, $G_X$ is the gain of the quadrature $X$, and $n_{\kappa,2}$ and $n_{\gamma}$ are defined in Eq.~\eqref{eq:n_kappas}. Assuming very low temperature operation of the loss channel, an overcoupling regime ($n_{\eta} \approx n_\gamma \approx 0$), a large HEMT gain $(G_\text{H}-1)/G_\text{H} \approx 1$, and knowing that the input variance to the KIPA is the vacuum (\textit{i.e.} $\braket{X^2_\text{1,in}} = \braket{X^2_\text{2,in}} = \braket{X^2_\text{b}} = \frac{1}{4}$), we can compare the variance of the measured output states with the KIPA turned on and off to obtain the squeezing factor $S$:
\begin{equation}
    S = \frac{\braket{X^2_\text{out}}|_\text{on}}{\braket{X^2_\text{out}}|_\text{off}} = \frac{G_X\eta\frac{1}{4} + (G_X-1)\eta n_{\kappa,2} + (1-\eta)\frac{1}{4} + n_\text{H}}{\frac{1}{4} + n_\text{H}}. \label{eq:squeezing}
\end{equation}
Using this relation, we can extract the squeezing gain $G_X$. Due to both attenuation mechanisms ($10 \log(\eta) = -4.5$~dB) and noise contribution from the HEMT ($n_\text{H} = n_\text{th,H}/2 = 3.25$~photons), we find that the maximum measurable vacuum squeezing \cite{Malnou_2019} for large gain ($|\xi| \to \kappa$, see Appendix~\ref{app:maxsqueezing}) is $S|_\text{min} = 0.9897$. This value is very close to our maximum measurement of squeezing $S$, as shown in Fig.~\ref{fig:squeeze}(c). We extract a maximum deamplification gain $G_X = -2.95$~dB using the model described by Eq.~\eqref{eq:squeezing}, which is close to the theoretical limit of $-3$~dB [see Appendix~\ref{app:maxsqueezing}].


\begin{figure*}
    \includegraphics[width=.96\linewidth]{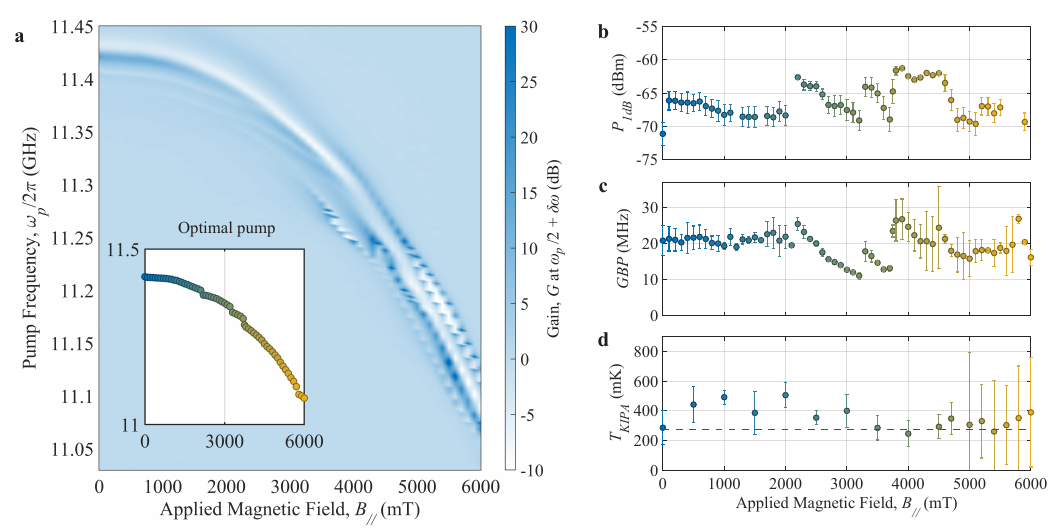}
	\caption{\textbf{Magnetic field characterization.} \textbf{(a)~Optimal pump frequency}. Non-degenerate gain as function of pump frequency ($\omega_p/2\pi$) and parallel magnetic field $B_\parallel$ obtained as explained in the main text. Inset: optimal pump frequency at each magnetic field, obtained as explained in the main text. \textbf{(b-d)~Magnetic field performance}. Measured saturation power $P_\text{1dB}$ (b), gain-bandwidth product $GBP$ (c) and KIPA noise temperature $T_\text{KIPA}$ (d) as a function of magnetic field. The red dashed line represents the half-photon quantum limit. The error bars take into account both fitting uncertainty and contributions from insertion loss asymmetry, as explained in the Appendix~\ref{app:tnoise}.}
	\label{fig:beffect}
\end{figure*}

\subsection*{\label{sec:magnetic}Effects of magnetic field}

We examined the impact of an in-plane magnetic field $B_\parallel$ on the KIPA performance by using a 6-1-1~Tesla magnet (three-axes magnet from American Magnets), integrated in our cryogenic system. The primary effect of an increasing magnetic fields consists of suppressing the superconducting gap $\Delta$, as described by the equation:
\begin{equation}
    \Delta(B) = \Delta_0 \sqrt{1- \bigg ( \frac{B}{B_C} \bigg) ^2},
\end{equation}
where $B_C$ represents the critical magnetic field, and $\Delta_0$ is the superconducting gap at zero field. To ensure the amplifier operates within a safe range of biasing currents, particularly when exposed to high magnetic fields, we conduct this study at a lower biasing current of $I_b = 50~\upmu$A. 
We performed a sweep of the in-plane magnetic field $B_{\parallel}$, ranging from 0 to 6 Tesla in increments of 100~mT. Figure~\ref{fig:beffect}(a) presents the KIPA gain at half the pump frequency plotted against the magnetic field. After stabilizing the magnetic field, we initiate the measurement with modest pumping power, gradually increasing it until identifying the optimal pump frequency that guarantees a 20~dB gain. This process was automated to efficiently find the optimal pumping conditions.


As illustrated in Fig.~\ref{fig:beffect}(a), the optimal pump frequency exhibits a trend with increasing magnetic field, resembling the behavior of the resonant frequency [see Appendix~\ref{app:magfieldfreqshift}], with the addition of occasional jumps. While we acknowledge that standing waves in the input, output and pump lines play a relevant role, we believe this effect may be enhanced by the different magnetic field dependence of wider and narrower sections of the SIF. In fact, as previously mentioned, biasing the device with a $dc$ currect leads to a faster redshift of the narrower sections of the KIPA, specifically the $Z_h$ and $Z_r$ segments, in comparison to the $Z_l$ segments. Because the low-impedance sections are inherently more susceptible to magnetic fields due to their wider cross-section, particularly in cases of slight magnetic field misalignment (estimated to be around $0.92 \pm 0.07$ degrees in our system), they undergo a stronger redshift under large magnetic fields. This ultimately results in the resonant frequency of the lower-impedance section of the SIF crossing the KIPA resonance, impacting the stability of the optimal pump frequency, which experience abrupt jumps in correspondence with those crossing points [see inset of Fig.~\ref{fig:beffect}(a)].

The evolution of the saturation power $P_\text{1dB}$ and the gain-bandwidth product $GBP$ are reported in Fig.~\ref{fig:beffect}(b-c). Gain and saturation power measurements were conducted at intervals of 100~mT. The KIPA exhibited robust performance up to 6 Tesla, maintaining a stable $P_\text{1dB} \approx -66 \pm 4$~dBm and a $GBP$ of roughly 20~MHz. However, in correspondence of the magnetic fields of $B_\parallel = 2.1, 3.3, 3.7$~Tesla, we observe abrupt jumps in both $P_\text{1dB}$ and $GBP$. We attribute these sudden changes in device performance to a combination of magnetic field-induced crossings and the presence of standing waves in the pump and signal lines, which lead to inaccuracies in estimating the optimal pump frequency of the amplifier [see inset of Fig.~\ref{fig:beffect}(a)].

Finally, we characterize the added noise of the KIPA $T_\text{KIPA}$ in magnetic field, following the same procedure previously described. Initially, we extract $T_\text{KIPA}$ at intervals of 500~mT, and above $B_\parallel = 4.5$~Tesla, with steps of approximately 200~mT. The estimated $T_\text{KIPA}$ is presented in Fig.~\ref{fig:beffect}(d). The associated error bars encompass both fitting uncertainties and contributions from insertion loss $\eta_{IL}$ [see Fig.~\ref{fig:noise}(a) and Appendix~\ref{app:tnoise}]. The amplifier consistently demonstrates noise performance approaching the quantum limit across the entire range of investigated magnetic fields, establishing it as a compelling device for applications in high magnetic field environments. This resilience to magnetic fields can be attributed to negligible magnetic field-induced loss when compared to the external coupling rate of the KIPA \cite{samkharadze_2016}.

\subsection*{\label{sec:temperature}Operation at high temperature}

\begin{figure}
    \centering
    \includegraphics[width=\linewidth]{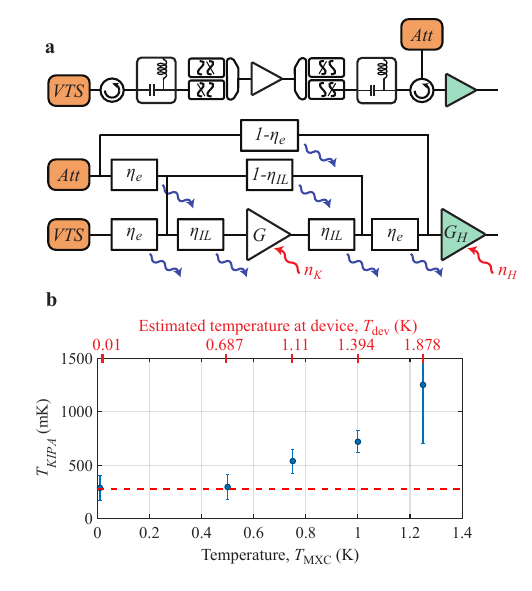}
	\caption{\textbf{Characterization at high temperature.} \textbf{(a) High temperature noise schematics}. Top: schematic representing the cryogenic components in series between the VTS and the HEMT, including  the contribution of the hot attenuator (corresponding to the 20~dB attenuator of the line Input 2 in Fig.~\ref{fig:expsetup}, anchored at MXC). Bottom: equivalent diagram of the modified cascaded transmission efficiencies and gains,leading to the added noise temperature equation \eqref{eq:tadd}. \textbf{(b) Noise temperature behaviour}. KIPA noise temperature as function of nominal ($T_\text{MXC}$, bottom $x$-axis in black) and estimated device ($T_\text{dev}$, top $x$-axis in red, see Appendix~\ref{app:hightempfreqshift}) operation temperature. The KIPA maintains quantum-limited performance up to $T_\text{MXC} > 500$~mK, corresponding to $T_\text{dev} > 687$~mK.}
	\label{fig:temperature}
\end{figure}

To conclude our study, we conduct measurements of the KIPA at higher temperature. An ideal superconducting parametric amplifier should not be significantly affected by changes in operating temperature, as long as superconductivity is preserved. The added noise photon number of the resonant KIPA \cite{Xu_2023b} is given by
\begin{equation}
    \bar{n} = \frac{\kappa}{\kappa + \gamma} n_{\text{bath}} + \frac{\gamma}{\kappa + \gamma} n_{\text{dev}}, \label{eq:radCool}
\end{equation}
where $\bar{n}$ represents the average photon population in the resonator, $\kappa$ [$\gamma$] is the coupling or external [internal] loss rate, and $n(T)~=~1/[\exp{(\hbar \omega_s / k_B T)}-1]$ is the average photon number estimated using the Bose-Einstein distribution, which is applicable to both the signal at temperature $T_{\text{bath}}$ and the device at temperature $T_{\text{dev}}$. In order to ensure proper quantum-limited operation, the radiative cooling condition needs to be satisfied, implying that $\bar{n} \approx n_{\text{bath}}$. 
This condition is typically met when the amplifier operates in the overcoupling regime, where $\kappa \gg \gamma$.
However, when the KIPA operates at temperatures that are a relevant fraction of the critical temperature, the quasi-particle population becomes noticeable as a dissipative channel, leading to an increase in the loss rate $\gamma$ \cite{frasca_2023}.

The operation temperature of the amplifier was controlled by gradually warming up the mixing chamber stage of the cryostat. This was performed by modifying the cooling power of the dilution unit, by adjusting the switch heater of the mixing chamber stage, effectively tuning the thermal contact between distillation and mixing chamber stages. The cryostat was stabilized at each designated temperature for two hours before proceeding with the measurement, to ensure minimal thermal drift. 


As the mixing chamber stage is warmed up, it becomes crucial to consider all the noise contributions of lossy cryogenic components. Adapted schematics depticting the cascaded transmission efficiencies and gains are presented in Fig.~\ref{fig:temperature}(a). The added noise, $T_\text{add}$, previously defined in Eq.~\eqref{eq:pout}, now becomes
\begin{align}
    T_\text{add} = & \frac{T_\text{HEMT}}{\eta G} + \frac{T_{Att}}{ \eta G} [1-\eta_e + \nonumber  \\ & \eta_e^2 (1-\eta_{IL}) + \eta^2 G] + T_\text{KIPA}. \label{eq:tadd}
\end{align}
Here, $T_{Att}$ is the attenuator temperature, assumed to be equal to the mixing chamber temperature due to the strong thermalization between the two. This adapted model allows us to extract the effective added noise of the KIPA ($T_\text{KIPA}$) when operating at higher temperatures.


In Figure \ref{fig:temperature}(b), we observe that $T_\text{KIPA}$ remains close to the quantum limit until the temperature of the mixing chamber, denoted as $T_\text{MXC}$, exceeds 500~mK. Beyond this point, we observe a gradual increase in $T_\text{KIPA}$. This evolution can be attributed to the presence of quasi-particles, which have a notable impact on the internal quality factor of an NbN resonator when the temperature exceeds 15\% of its critical temperature ($T_C$) \cite{frasca_2023}.


However, when considering the resonator frequency shift [as shown in Fig.~\ref{fig:TFit}], we can estimate a higher operational temperature, denoted as $T_\text{dev}$, for the device hosted in the FSE as compared to the mixing chamber temperature ($T_\text{MXC}$). This temperature difference may arise due to significant temperature gradients between the mixing chamber and the FSE system, particularly when reducing the cooling power at the mixing chamber. Assuming that the resonator frequency shift provides a more accurate estimate of the effective device temperature compared to the thermometer installed at the mixing chamber ($T_\text{MXC}$), we can infer that the device exhibits approximately quantum-limited noise performance up to an operational temperature of approximately $1.11$~K.
To further validate the noise performance of the KIPA, we conducted characterizations by mounting the device on the still plate of our cryogenic system, achieving a stable operational temperature of approximately $850$~mK. Detailed results of these measurements are presented in Appendix~\ref{app:still}, providing confirmation of quantum-limited noise performance at this temperature.

\subsection*{\label{sec:conclusion}Conclusions}


In conclusion, our study offers a comprehensive exploration of the capabilities and performance of a Kinetic Inductance Parametric Amplifier (KIPA) based on high kinetic inductance superconducting NbN thin films.
This investigation highlights the outstanding performance of the NbN KIPA, particularly when operating near its critical conditions.
The KIPA consistently approaches the quantum noise limit, maintaining a stable maximum gain, gain-bandwidth product, and saturation power even under the influence of a 6 Tesla in-plane magnetic field and at temperatures up to 850~mK.
Furthermore, we have demonstrated its ability to generate vacuum squeezed states, confirming the KIPA's quantum-limited noise performance.
One of the remarkable features of the presented NbN KIPA is its wide frequency tunability range of approximately 400~MHz ($\sim 7$\% of the KIPA resonant frequency), deriving from the high quality and uniformity of the deposited superconducting NbN film. 

The KIPA's primary strength lies in its potential for integration into a wide range of experimental settings. Its ability to operate at or near the quantum noise limit up to 850~mK, combined with its compatibility with large magnetic fields, positions it as an ideal candidate for integration with spin-based quantum technology. Notably, quantum dot-based spin qubits have recently demonstrated functionality at temperatures as high as 4 K \cite{camenzind_2022}, opening exciting opportunities for direct on-chip integration.




In summary, the KIPA's exceptional noise performance, its capacity to handle higher saturation powers, its partial frequency tunability, and its resilience to higher temperatures and magnetic fields make it a compelling alternative to conventional Josephson-based parametric amplifiers. 
These features collectively establish the KIPA as a powerful tool with the potential to advance research across diverse domains of quantum science and technology. 
Its versatility and robustness in extreme temperature and magnetic field conditions promise to drive innovation and accelerate discoveries in fundamental quantum research and applications.

\section*{Contributions}

S.F. and P.S. conceived the experiments. S.F. developed the fabrication recipes and optimized the deposition and characterization techniques. S.F. fabricated the device. S.F. and C.R. performed measurements. S.F. and G.B. analyzed the data. S.F. wrote the manuscript with inputs from all authors. P.S. supervised the work.

\section*{Acknowledgements}

This research was funded by Swiss National Science Foundation through the project grants NCCR SNF 51AU40-1180604 and SNF project 200021-200418. S.F. acknowledges the support of SNF Spark project 221051.

\appendix

\begin{figure}	
    \centering
    \includegraphics[width=\linewidth]{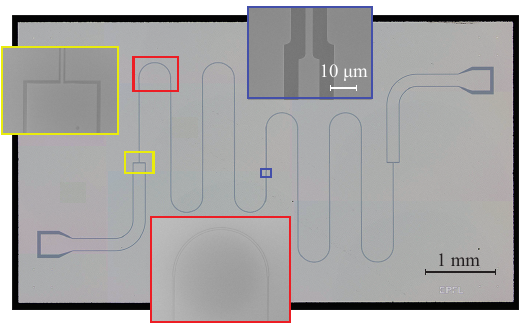}
    \caption{\textbf{Device Micrograph}. Full size optical micrograph of the KIPA. In the yellow frame: the $50~\Omega$-to-$Z_l$ transition; in the red frame: the bending radius of the device, chosen to be much larger than the center conductors to ensure good impedance control; in the blue frame: $Z_l$-to-$Z_h$ transition. The small (length of 3~$\upmu$m) hyperbolic section is to minimize potential $dc$ current crowding effects.}
    \label{fig:device}
\end{figure}

\section{\label{app:materials}Materials}

\begin{figure*}	
    \centering
    \includegraphics[width=.96\linewidth]{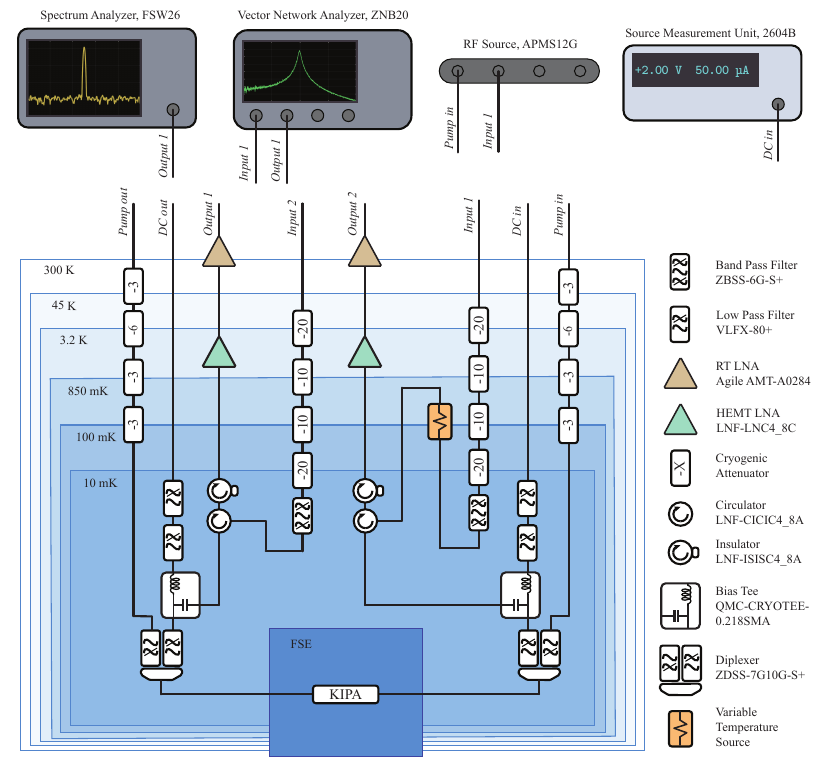}
    \caption{\textbf{Experimental Setup}. Detailed schematic illustrating the experimental setup used for noise and gain measurements. The symmetric setup design allows to detect any device asymmetry accurately. The input lines are attenuated with 60~dB cryogenic attenuators distributed across the several stages of the dilution refrigerator. One of the input lines is routed to the ``cold plate'' stage (typically at a temperature $T_\text{CP}=100$~mK) through a 20~dB variable temperature source (VTS) \cite{Simbierowicz_2021}. The $dc$ current is delivered through non-attenuated and filtered SMA lines, while the device pump is routed through 15~dB attenuated lines. Both the pump tone and $dc$ current are combined with the signal tone at base temperature using a bias-tee and a diplexer. The output line is connected through circulators to a cryogenic HEMT low-noise amplifier anchored at the 4~K stage and to a room temperature amplifier. An additional 10~dB [18.5~dB] attenuation on the input lines, respectively for signal at frequency $\sim$6~GHz [$\sim$12~GHz], needs to be taken into account, coming from the intrinsic coaxial lines losses.}
    \label{fig:expsetup}
\end{figure*}

\subsection{\label{app:fab}Device fabrication}

The device fabrication starts with a 2~minutes substrate immersion in a 40\% HF solution, effectively eliminating native oxide and potential contaminants from the surface of an intrinsic 2-inch Si wafer, with high--resistivity ($\rho \ge 10~k\Omega$cm) and $\braket{100}$ crystalline orientation. There follows an NbN films bias sputtering \cite{dane_bias_2017} at room temperature in a Kenosistec RF sputtering system, with a thickness of 13~nm. Bias sputtering offers the advantage of generating polycrystalline films, which typically exhibit enhanced device yield. Film deposition occurs at 5~$\upmu$bar pressure, with 8\% N$_2$ in Ar atmosphere, resulting in films with room-temperature sheet resistance of 250~$\Omega/\square$, corresponding to a sheet inductance of approximately 100~pH/$\square$ \cite{frasca_2023}. After subsequent deposition of Ti/Pt alignment markers by optical lift-off process, and dehydration step at 150$^\circ$C for 5~minutes, 50~nm-thick ZEP-502A positive e-beam resist, diluted at 33\% in anisole, is spin coated at 4000~rpm on the wafer, and baked at 150$^\circ$C for 5~minutes. With an electron beam lithography (Raith EBPG5000+ at 100~keV) step, the device is patterned on the resist through development in amyl acetate for 1~minute, followed by rinsing in a solution 9:1~MiBK:IPA. The pattern in the resist mask is then transferred to the NbN using CF$_\text{4}$/Ar mixture and reactive ion etching with a power of 15~W and pressure of 2~mTorr for 5 minutes. The resist is then stripped with Microposit remover~1165 heated at 70$^\circ$C. Finally, the wafer is coated with 1.5~$\upmu$m AZ~ECI~3007 positive photolithography resist for devices protection and diced. A typical device is represented in Fig.~\ref{fig:device}.

\section{\label{app:methods}Methods}

\subsection{\label{app:setup}Experimental Setup}

All the measurements are conducted in a BlueFors system equipped with fast sample exchange (FSE) option. Signal generation and readout are performed with either a 2-port VNA (Rohde \& Schwarz, ZNB20) or via a RF source (Anapico, APMS12G) and a Spectrum Analyzer (Rohde \& Schwarz, FSW26) respectively. The pump tone is handled by the same RF source while the $dc$ bias is provided by a Source Measurement Unit (Keithley, 2604B).

The setup allows to measure the KIPA in transmission in two different ways as depicted in Fig.~\ref{fig:expsetup}. In the first configuration, the input signal (see \textit{Input 1}) generated by the RF source enters the cryostat and goes though a set of cryogenic attenuators adding up to 60~dB, and a band pass filter (MiniCircuit, ZBSS-6G-S+) before being rerouted to the cold plate stage, $T_\text{CP} \approx 100$~mK, where the signal is fed though a Variable Temperature Source (BlueFors, VTS). After a first circulator (Low Noise Factory, LNF-CICIC4\_8A), the input signal is then combined with the low pass filtered (MiniCircuit, VLFX-80+) $dc$ signal (from line \textit{DC In}) via a bias tee (Quantum Microwave, QMC-CRYOTEE-0.218SMA) and then combined with the incoming $15$~dB attenuated pump (\textit{Pump In}) via a Diplexer (MiniCircuit, ZDSS-7G10G-S+) before entering the KIPA. After the signal passes though the KIPA, the pump and $dc$ are separated using a similar scheme and extinguished at room temperature. The output signal follows though a circulator (Low Noise Factory, LNF-ICICC4\_8A) and an isolator (Low Noise Factory, LNF-ISISC4\_8A) before being further amplified by a HEMT (Low Noise Factory, LNF-LNC4\_8C) at the 4K stage and a Low Noise Amplifier (Agile, AMT-A0284) at RT. The signal is collected either by the second port of the VNA or the Spectrum Analyser connected at \textit{Output 1}.

In a symmetrical way, the setup can be inverted so that the signal enters via \textit{Input 2} and out though \textit{Output 2}. Changing the direction of pump and bias did not affect the operation of the device, which suggest good symmetry between the two ports of the amplifier.

\subsection{\label{app:magsetup}Magnetic field measurements}

Magnetic field sweeps are carried out using a Integrated cryogen-free superconducting magnet from American Magnetics Inc. (Cryogen-Free MAxes Vector Magnet), capable of applying magnetic fields up to 6 Tesla in the $z$ direction (in plane field) and up to 1 Tesla in the $x-y$ plane. The magnet is integrated in the Bluefors cryostat at the 4K stage. During measurements, the device is placed within the FSE at the location where the magnetic field is maximum. Each magnet lead is powered by a Model 430 power supply from the same manufacturer, enabling the generation of a ramping current in both circuit directions to achieve both positive and negative field values. A persistent switch can be deactivated to keep the supercurrent in the magnet coil without providing power. However, to ensure stability of the magnetic field, in the presented experiments the switch is always kept on. The magnetic field reading is obtained by measuring the current in the leads, with a resolution of a tenth of mT. 

The magnetic field is adjusted to a ramp rate of 10 mT/min, with a series of measurements (optimal pump search, gain, saturation power) automatically initiated every 100~mT. Noise measurements are conducted at intervals of 500~mT up to an in-plane magnetic field strength of 4.5~T. Beyond this point, noise measurements are done every 200~mT to ensure a more precise study. 

\subsection{\label{app:coupling}Coupling of the KIPA}

Following the description of a quarter-wave transformer, we can estimate the coupling quality factor of the KIPA. In general, the input impedance seen from a load is defined by \cite{pozar_2012}:
\begin{equation}
    Z_{\text{in}} = Z_f \frac{Z_0 + i Z_f \tan{(\beta l)}}{Z_f + i Z_0 \tan{(\beta l)}},
\end{equation}
where $Z_f$ is the impedance of the section of length $l$ and complex propagation constant $\beta = 2\pi/\lambda$. In the limit where the length of the section is either $\lambda/2$ or $\lambda/4$, we get
\begin{equation}
    \lim_{l \to \lambda/4} Z_{\text{in}} = \frac{Z_f^2}{Z_0}, \quad \lim_{l \to \lambda/2} Z_{\text{in}} = Z_0.
\end{equation}
In our design, the KIPA decoupling is made from two stepped-impedance filters (SIF), one on each side, which are designed as a sequence of high and low impedance $\lambda/4$ sections, with a central resonator of length $\lambda/2$ and impedance $Z_r$. This configuration is the microwave equivalent of a Fabry-P\'erot cavity. The two mirrors have $n_l = 6$ sections of $Z_l$ and $n_h = 5$ sections of $Z_h$, hence the impedance seen by the central resonator is:
\begin{equation}
    Z_{\text{eff}} = \frac{Z_l^{2 n_l}}{Z_h^{2 n_h}} \frac{1}{Z_0},
\end{equation}
with $Z_0 = 50~\Omega$ is the lines impedance. The associated coupling quality factor for a series circuit reduces to \cite{pozar_2012}:
\begin{equation}
    Q_c = \frac{\omega_r L_r}{Z_{\text{eff}}} = \frac{4 Z_r}{\pi Z_{\text{eff}}}.
\end{equation}
Hence, for our KIPA impedance of $Z_r = 900~\Omega$, the coupling quality factor $Q_c = 290$, which leads to a resonator coupling $\kappa/2\pi = \omega_r/ 2\pi Q_c = 19.8$~MHz.

\begin{figure}
    \centering
    \includegraphics[width=\linewidth]{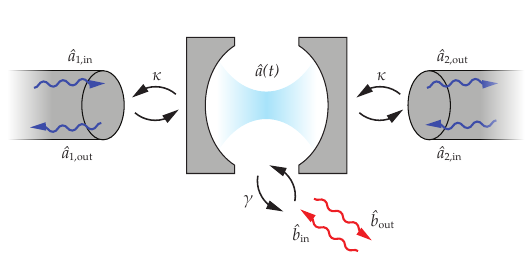}
	\caption{\textbf{Quantum optics cavity representation}. Schematic representation of the microwave cavity described by the Hamiltonian $H_\text{KIPA}$ [see Eq.~\eqref{eq:Hkipa}]. The cavity is coupled to two microwave ports in a transmission configuration.}
	\label{fig:scattering}
\end{figure}

\subsection{\label{app:iotheory}Scattering parameters estimation}

To estimate the transmission spectrum of the resonator, we treat it as a cavity with input and output ports [see Fig.~\ref{fig:scattering}]. The evolution of the intra-cavity field $\hat{a}(t)$ is given by the quantum Langevin equation:
\begin{align}
    \frac{\partial \hat{a}(t)}{\partial t} = & \frac{[\hat{a}(t),H_{\text{KIPA}}]}{i\hbar} - \bar{\kappa} \hat{a}(t) + \sqrt{\kappa_1} \hat{a}_{\text{1,in}}(t) +\nonumber \\ & \sqrt{\kappa_2} \hat{a}_{\text{2,in}}(t) + \sqrt{\gamma}\hat{b}_{\text{in}}(t)
\end{align}
with $\bar{\kappa} = (\kappa_1 + \kappa_2 + \gamma)/2$, $\hat{a}(t)$ is the resonator field, and $\hat{a}_{i,\text{in}}(t)$ [$\hat{b}_{\text{in}}(t)$] is the input field coupled to the input or output line [to the environment] with a coupling rate $\kappa_i$ [$\gamma$]. The input-output relations are:
\begin{equation}
     \hat{a}_{1,\text{out}}(t) + \hat{a}_{1,\text{in}}(t) = \sqrt{\kappa_1}\hat{a}(t),
\end{equation}
\begin{equation}\label{eq:input-output k2}
     \hat{a}_{2,\text{out}}(t) + \hat{a}_{2,\text{in}}(t) = \sqrt{\kappa_2}\hat{a}(t),
\end{equation}
\begin{equation}
     \hat{b}_{\text{out}}(t) + \hat{b}_{\text{in}}(t) = \sqrt{\gamma}\hat{a}(t).
\end{equation}

Following the same formalism as in \cite{parker_2022, chen_2022}, we make use of the input-output relation of Eq.~\eqref{eq:input-output k2} to link the transmitted field $\hat{a}_{2,\text{out}}$ to the input fields. Assuming that the two ports have the same coupling $\kappa_{1,2} = \kappa$, we find 
\begin{equation}\label{eq:input-ouput-gain}
\begin{split}
    \hat{a}_{2,\text{out}}(\omega) =& g_s(\omega) \hat{a}_{1,\text{in}}(\omega) +g_i(\omega) \hat{a}_{1,\text{in}}^\dag(\omega)\\ & +\sqrt{\frac{\gamma}{\kappa}} \left( g_s(\omega)\hat{b}_{\text{in}}(\omega) + g_i(\omega)\hat{b}_{\text{in}}^\dag(\omega) \right) \\ & + \left( [g_s(\omega)-1] \hat{a}_{2,\text{in}}(\omega) + g_i(\omega)\hat{a}_{2,\text{in}}(\omega) \right), 
\end{split}
\end{equation}
where the signal $g_s(\omega)$ and idler $g_i(\omega)$ gains correspond to:
\begin{equation}
    g_s(\omega) = \frac{\kappa \bar{\kappa} - i \kappa (\Delta + \omega)}{\Delta^2 + (\bar{\kappa} - i\omega)^2 - |\xi|^2}, \label{eq:signalG}
\end{equation}
\begin{equation}
    g_i(\omega) = \frac{-i \xi \kappa}{\Delta^2 + (\bar{\kappa} - i\omega)^2 - |\xi|^2}.\label{eq:idlerG}
\end{equation}
The Eq.~\eqref{eq:input-ouput-gain} can be combined with its Hermitian conjugate to get the input-ouput relations for the two quadratures
\begin{equation}
\begin{split}
    \hat{X}_{2,\text{out}} =& \frac{1}{2} \left(  \hat{a}_{2,\text{out}}^\dag+ \hat{a}_{2,\text{out}} \right)\\
    =&\frac{1}{2} [\epsilon \hat{a}_{1,\text{in}}+\epsilon^*\hat{a}_{1,\text{in}}^\dag +\sqrt{\frac{\gamma}{\kappa}} (\epsilon \hat{b}_{\text{in}} 
    + \epsilon^*\hat{b}_{\text{in}}^\dag )\\+&(\epsilon-1)\hat{a}_{2,\text{in}} +(\epsilon^*-1)\hat{a}_{2,\text{in}}^\dag ],
\end{split}
\end{equation}
\begin{equation}\label{eq:y_quad}
\begin{split}
    \hat{Y}_{2,\text{out}} =& \frac{i}{2} \left(  \hat{a}_{2,\text{out}}^\dag- \hat{a}_{2,\text{out}} \right)\\
    =&\frac{1}{2} [\epsilon'^* \hat{a}_{1,\text{in}}^\dag -\epsilon' \hat{a}_{1,\text{in}}+\sqrt{\frac{\gamma}{\kappa}} (\epsilon'^* \hat{b}_{\text{in}}^\dag 
    - \epsilon'\hat{b}_{\text{in}} )\\+&(\epsilon'^*-1)\hat{a}_{2,\text{in}}^\dag +(\epsilon'-1)\hat{a}_{2,\text{in}} ],
    \end{split}
\end{equation}
where $\epsilon=g_s+g_i^*$ and $\epsilon'=g_s-g_i^*$. These two equations can be combined with Eq.~\eqref{eq:input-ouput-gain} to obtain 
\begin{equation}
\begin{split}
    \begin{pmatrix}
\hat{X}_{2,\text{out}} \\ 
\hat{Y}_{2,\text{out}}
\end{pmatrix}= & G_m\begin{pmatrix}
\hat{X}_{1,\text{in}} \\ 
\hat{Y}_{1,\text{in}}
\end{pmatrix}+(G_m-1)\begin{pmatrix}
\hat{X}_{2,\text{in}} \\ 
\hat{Y}_{2,\text{in}}
\end{pmatrix}\\+&\sqrt{\frac{\gamma}{\kappa}}G_m\begin{pmatrix}
\hat{X}_{b,\text{in}} \\ 
\hat{Y}_{b,\text{in}}
\end{pmatrix} ,
\end{split}
\end{equation}
where $\hat{X}_{1/2/b,\text{in}}$ and $\hat{Y}_{1/2/b,\text{in}}$ are the quadratures of $\hat{b}_{\text{in}}/\hat{a}_{1/2,\text{in}}$ and
\begin{equation}
    G_m=\begin{pmatrix}
\text{Re}(\epsilon) & -\text{Im}(\epsilon) \\ 
 \text{Im}(\epsilon') & \text{Re}(\epsilon')
\end{pmatrix}.
\end{equation}
Assuming that all quadratures of the input fields are uncorrelated and that we are operating at the point $\psi_p=0$ and $\Delta=0$, the variance $\braket{\hat{X}_{2,\text{out}} ^2}$ is simply given by 
\begin{equation}
    \braket{\hat{X}_{2,\text{out}}^2}= G_X \braket{\hat{X}_{1,\text{in}}^2}+(G_X-1)n_{\kappa,2}+G_X n_{\gamma}, \label{eq:Xout2}
\end{equation}
where $G_X=\text{Re}(\epsilon)$ and,
\begin{align}
    &n_{\kappa,2} = \frac{\sqrt{G_X}-1}{\sqrt{G_X}+1} \braket{X^2_\text{2,in}}\nonumber \\ & n_{\gamma} = \frac{\gamma}{\kappa} \braket{X^2_\text{b,in}}. 
    \label{eq:n_kappas}
\end{align}

\subsection{\label{app:maxsqueezing}Squeezing limit in transmission architecture}

When operating the KIPA in transmission, the maximum achievable vacuum squeezing is limited to $-3$~dB. This can be explained by considering that a KIPA operated in transmission experiences two loss channels: its intrinsic loss rate $\gamma$ and the second port $\kappa$.
The contribution of the second port, through $n_{\kappa,2}$, affects the output variance $\braket{\hat{X}_{2,\text{out}}^2}$ as described in Eq.~\eqref{eq:Xout2}.

This interpretation is further confirmed by looking at the phase-sensitive gain equation [see Eq.~\eqref{eq:degenerate}]. Assuming that the pump is effectively operating at twice the drifted resonator frequency, \textit{i.e.} with $\Delta = 0$, Eq.~\eqref{eq:degenerate} can be written as
\begin{equation}
    |g(\psi_p)| = \bigg | \frac{\kappa \bar{\kappa} + i \kappa |\xi| e^{-i\psi_p}}{\bar{\kappa}^2 - |\xi|^2} \bigg |, \label{eq:phaseSensG}
\end{equation}
where $\bar{\kappa} = (2 \kappa + \gamma)/2$. To ensure that idler and signal modes in the cavity interact destructively, we set $\psi_p = 0$, which leads to
\begin{equation}
    |g(0)| = \bigg | \frac{\kappa \bar{\kappa} - \kappa |\xi|}{\bar{\kappa}^2 - |\xi|^2}  \bigg |  = \bigg | \frac{\kappa}{(\bar{\kappa} +|\xi|)} \bigg |. \label{eq:squeezing2}
\end{equation}
From this we get that, for low pumping, $|\xi| \to 0$ and $|g| \to 1$, while for very large gain, $|\xi| \to \kappa$ and $|g| \to 1/2$.

\begin{figure}
    \centering
    \includegraphics[width=\linewidth]{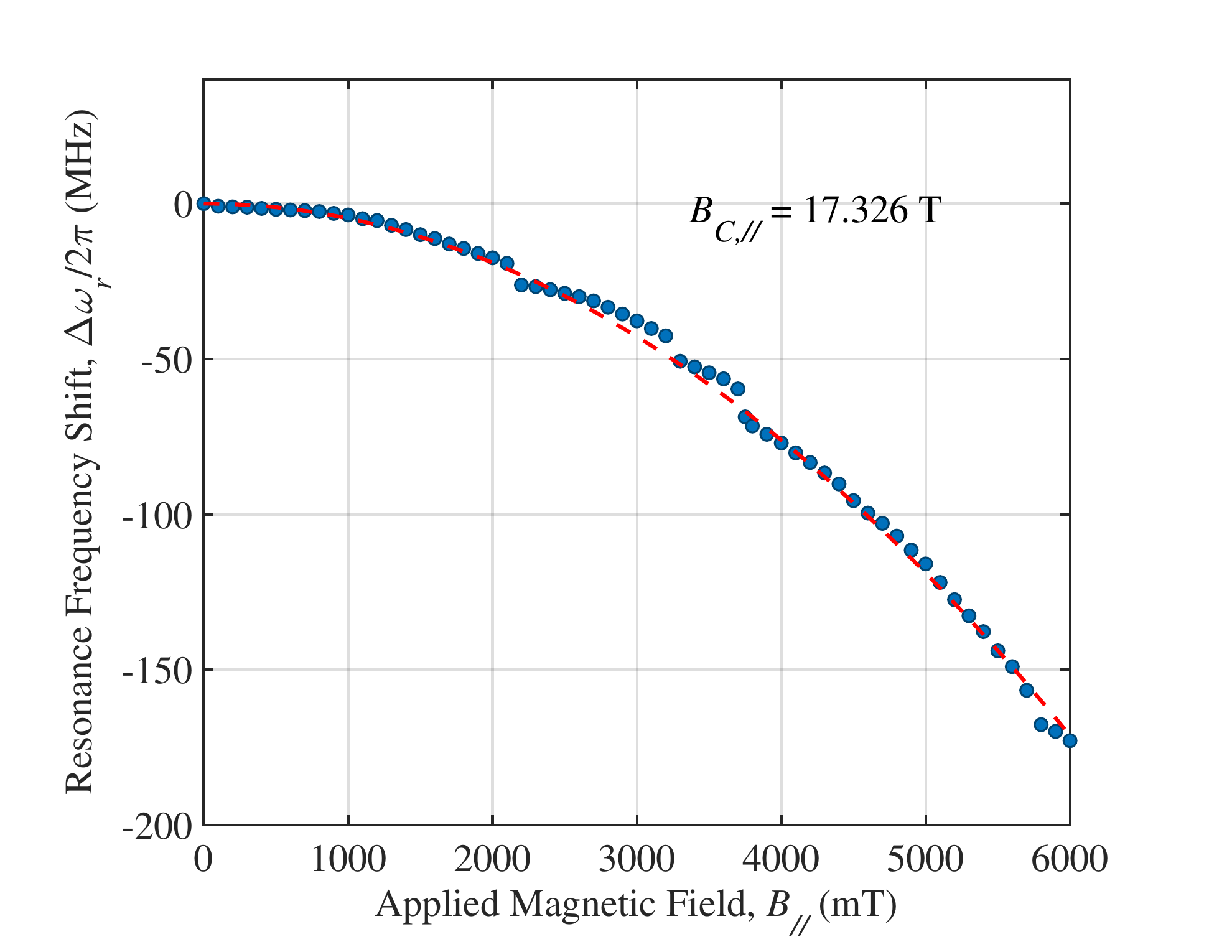}
	\caption{\textbf{Critical magnetic field}. Frequency shift $\Delta \omega_r$ of the KIPA induced by the magnetic field $B_\parallel$, with $\omega_r = \omega_p/2$. The red dashed line illustrates the fitting results using Eq.~\eqref{eq:BCfield}. The analysis yields a critical magnetic field $B_{C,\parallel}$ of 17.32~Tesla.} 
	\label{fig:BFit}
\end{figure}

On the other hand, operating the device in reflection allows to measure a much larger squeezing as long as the overcoupling condition is satisfied. In fact, as per Ref.~\cite{parker_2022}, the maximum achievable deamplification is given by
\begin{equation}
    |g(0)| = \bigg | \frac{\kappa \tilde{\kappa} - \kappa |\xi|}{\tilde{\kappa}^2 - |\xi|^2}  -1 \bigg |  = \bigg | \frac{\kappa}{(\tilde{\kappa} +|\xi|)} -1 \bigg |. \label{eq:squeezing3}
\end{equation}
where $\tilde{\kappa} = (\kappa + \gamma)/2$. In the large gain condition for a KIPA operated in reflection configuration, $|\xi| \to \kappa/2$ and $|g| \to 0$, leading to a much larger upper limit for measurable squeezing.

\subsection{\label{app:magfieldfreqshift}Critical magnetic field}

We estimate the critical in-plane magnetic field $B_{C,\parallel}$ of our KIPA by following the shift in its resonant frequency $\Delta \omega_r = \omega_r(B) - \omega_{r,0}$ as function of applied magnetic field $B_\parallel$. Here, $\omega_{r,0}$ represents the KIPA resonant frequency at zero field. We assume that $\omega_r(B)$ is half of the optimal pump frequency, as described in Fig.~\ref{fig:beffect}. By fitting $\Delta \omega_r$ [see Fig.~\ref{fig:BFit}] with respect to the magnetic field, following the equation \cite{Khalifa_2023}:

\begin{equation}
\frac{\Delta \omega_r}{\omega_r} = - \frac{\pi}{48}\frac{D e^2 t^2}{\hbar k_B T_C} \bigg[1 + \theta_B^2 \bigg(\frac{w_r}{t} \bigg)^2 \bigg] B_{\parallel}^2, \label{eq:BCfield}
\end{equation}
we extract a critical parallel magnetic field of $B_{C,\parallel} = 17.32$~Tesla. Here, $D = 0.5$~cm$^2$s$^{-1}$ is the electron diffusion coefficient, $T_C = 5.6$~K is the critical temperature of the NbN thin film superconductor \cite{frasca_2023}, $w_r$ denotes the center resonator width, $t$ its thickness, and $\theta_B$ is the magnetic field misalignment angle, which is estimated to be $0.92 \pm 0.07$~degrees.

\begin{figure}
    \centering
    \includegraphics[width=\linewidth]{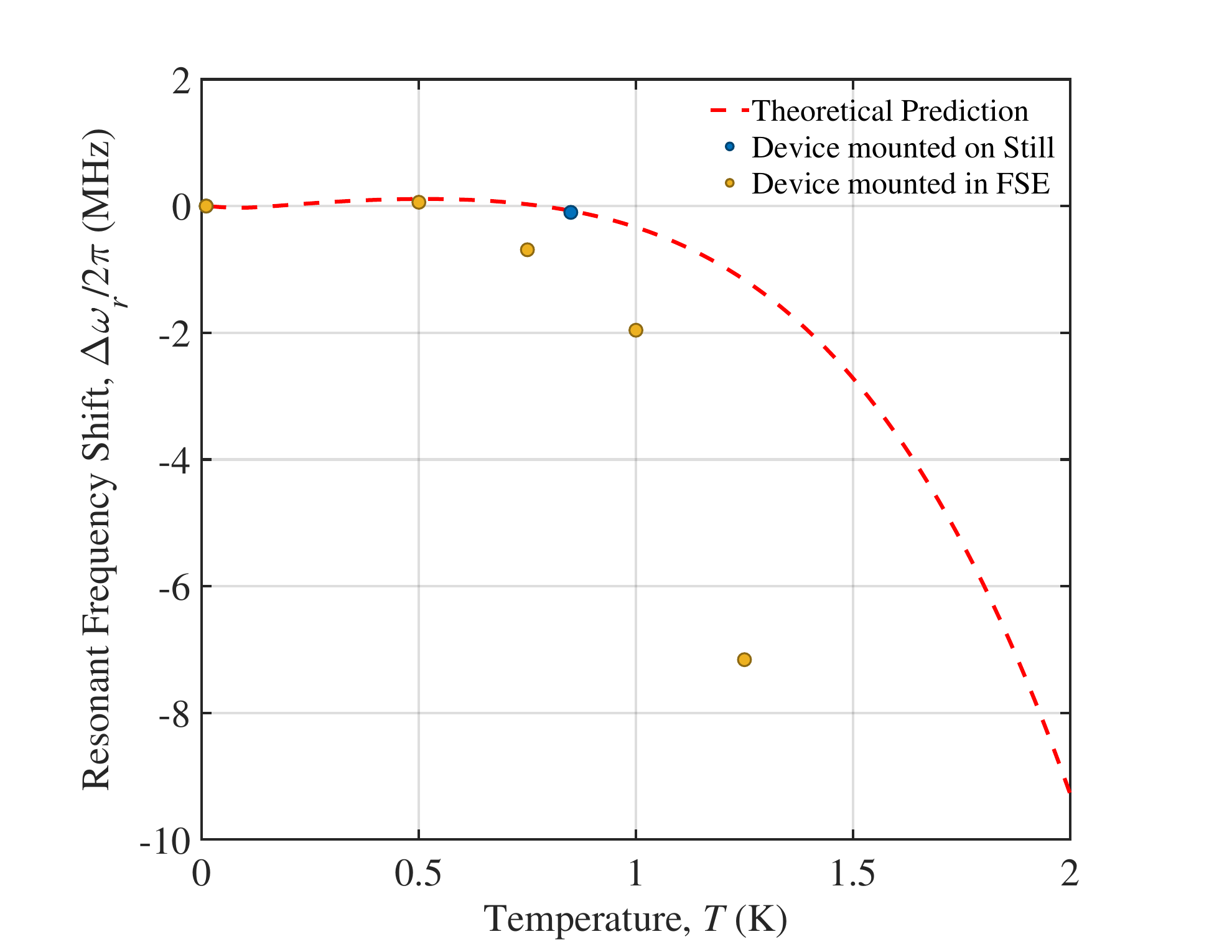}
	\caption{\textbf{Temperature induced frequency shift}. Resonant frequency shift, $\Delta \omega_r$ measured at various temperatures. Yellow datapoints correspond to measurements performed in the FSE, while the blue point represents the KIPA frequency shift operating at cryostat's still plate. The red dashed line illustrates the predicted shift in resonant frequency for the KIPA model based on Eq.~\eqref{eq:df_final}. The findings suggest a substantial temperature gradient within the system, with the FSE temperature estimated to be significantly higher than that of the mixing chamber.}
	\label{fig:TFit}
\end{figure}

\subsection{\label{app:hightempfreqshift}High temperature frequency shift}

In order to estimate the effective operating temperature of the KIPA ($T_\text{dev}$) when installed in the FSE and during the temperature sweep [see Fig.~\ref{fig:temperature}(b)], we refer to a previous study conducted in Ref.~\cite{frasca_2023}. 
The device resonant frequency shift $\Delta \omega_r$ as function of the temperature is described by the following equation:
\begin{align}
    \frac{\Delta \omega_r}{\omega_r} = &~\frac{F\delta^0_\text{TLS}}{\pi} \bigg( \text{Re} \bigg\{ \Psi \bigg( \frac{1}{2}+\frac{h f_r}{2 i \pi k_B T} \bigg) \bigg\} + \nonumber 
    \\ & - \ln{\frac{h f_r}{2 \pi k_B T}} \bigg) - \alpha \frac{\Delta L_\text{k}}{L_\text{k}}. \label{eq:df_final}
\end{align}
Here, $F\delta^0_\text{TLS}$ represents the contribution from two-level fluctuators, $\Psi$ is the digamma function \cite{Pappas_2011}, $\alpha$ is the fraction of kinetic inductance contributing to the overall inductance of the device, $\frac{\Delta L_\text{k}}{L_\text{k}}$ represents the change in the kinetic inductance, $L_\text{k}(T) = \mu_0 \lambda^2(T) (l/wd)$, and $\lambda(T)$ is the temperature-dependent London penetration depth.
As in Ref.~\cite{frasca_2023}, we assume that the film behaves as a superconductor in the dirty limit.
It follows that
\begin{equation}
    \frac{\Delta L_\text{k}}{L_\text{k}} = \frac{\lambda^2 (T) - \lambda^2 (0)}{\lambda^2 (0)} \propto \bigg( \frac{k_B T}{\Delta} \bigg)^4,
\end{equation}
where $\Delta = 1.764~k_BT_C$ is the superconducting gap and $T_C = 5.6$~K is the critical temperature of the superconducting film, given that $[\lambda(T)/\lambda(0) - 1] \propto (T/T_C)^2$ \cite{Eliashberg_1991-lh,shoji_1992,lee_1993} for $T \le 0.4~T_C$ for a superconductor in the dirty limit. In Fig.~\ref{fig:TFit} we present in blue the $\Delta \omega_r$ measured by thermally anchoring the KIPA to the Still stage $T_\text{dev} = T_\text{Still} = 850$~mK [see Appendix~\ref{app:still}]; instead the yellow data points represent $\Delta \omega_r$ measured assuming $T_\text{dev} = T_\text{MXC}$ during the temperature sweep reported in Fig.~\ref{fig:temperature}. 
The red dashed line corresponds to the theoretical prediction of $\Delta \omega_r(T)$ according to Eq.~\eqref{eq:df_final}.
While the blue data point closely aligns with the predicted value, this is not the case for operations in the FSE (yellow points). This temperature difference can arise from suboptimal cooling power operation conditions, potentially resulting in significant temperature gradients within the mixing chamber stage. Utilizing the measured frequency shifts, we calculate the effective $T_\text{dev}$ in corresponcence of different mixing chmaber temperatures (highlighted in red in the top $x$-axis in Fig.~\ref{fig:temperature}).

\subsection{\label{app:tnoise}Noise temperature measurements: $T_\text{in}$}

To accurately calibrate the noise introduced by the variable temperature source (VTS) at the input of the KIPA, it is essential to precisely quantify the attenuation between the VTS and the KIPA. The VTS connects to a port of the circulator at the mixing chamber through NbTi coaxial lines, ensuring no thermalization nor attenuation of the input power. Subsequently, after passing through the bias tee necessary for device DC bias, and the diplexer used to introduce the pump tone, the signal is injected into the KIPA. 
The temperature of the VTS ($T_{\text{VTS}}$) is controlled, allowing for the calculation of the output noise temperature $T_\text{out}$, as detailed in \cite{Simbierowicz_2021}:
\begin{align}
    T_\text{out} = & \frac{\hbar \omega_s}{2 k_B} \coth{\frac{\hbar \omega_s}{2 k_B T_{\text{VTS}}}} +  \nonumber  \\ & \frac{G_{\text{conv}}}{G} \frac{\hbar \omega_i}{2 k_B} \coth{\frac{\hbar \omega_i}{2 k_B T_{\text{VTS}}}}. \label{eq:tout}
\end{align}
Here, $G_{\text{conv}}$ is the conversion ratio of signal and idler, taken as $G-1$ for simplicity, and $G$ denotes the KIPA gain. Having accurately modeled the losses attributed to the attenuation of lines and cryogenic components at the MXC stage, we can assess the power reaching the amplifier using the expression:
\begin{equation}
    T_\text{in} = \eta T_\text{out}. \label{tin}
\end{equation}
Here, $\eta$ is defined as the product of $\eta_{e}$ and $\eta_{IL}$, where $\eta_{e}$ represents the total loss from the VTS to the amplifier due to electronics, given by $\eta_{e} = \eta_{CC} \times \eta_{coax} \times \eta_{BT} \times \eta_{DX}$. The individual components of $\eta_{e}$ include losses from the LNF-ICIC4\_8C circulator [$10 \log(\eta_{CC}) = -0.2$~dB], losses from approximately 0.7 m of SMA coaxial cables [$10 \log(\eta_{coax}) = -0.6$~dB], losses from the QMC-CRYOTEE-0.218SMA bias tee [$10 \log(\eta_{BT}) = -0.2$~dB], and losses from the ZDSS-7G10G-S+ diplexer [$10 \log(\eta_{DX}) = -0.25$~dB]. Additionally, $\eta_{IL}$ represents the frequency-dependent insertion loss of the amplifier.

We assume an equal division of the total transmission insertion loss of the amplifier between the two ports. Any potential asymmetry that might introduce additional error to the input noise power estimation is duly considered by the error bars.

\begin{figure*}
    \centering
    \includegraphics[width=.96\linewidth]{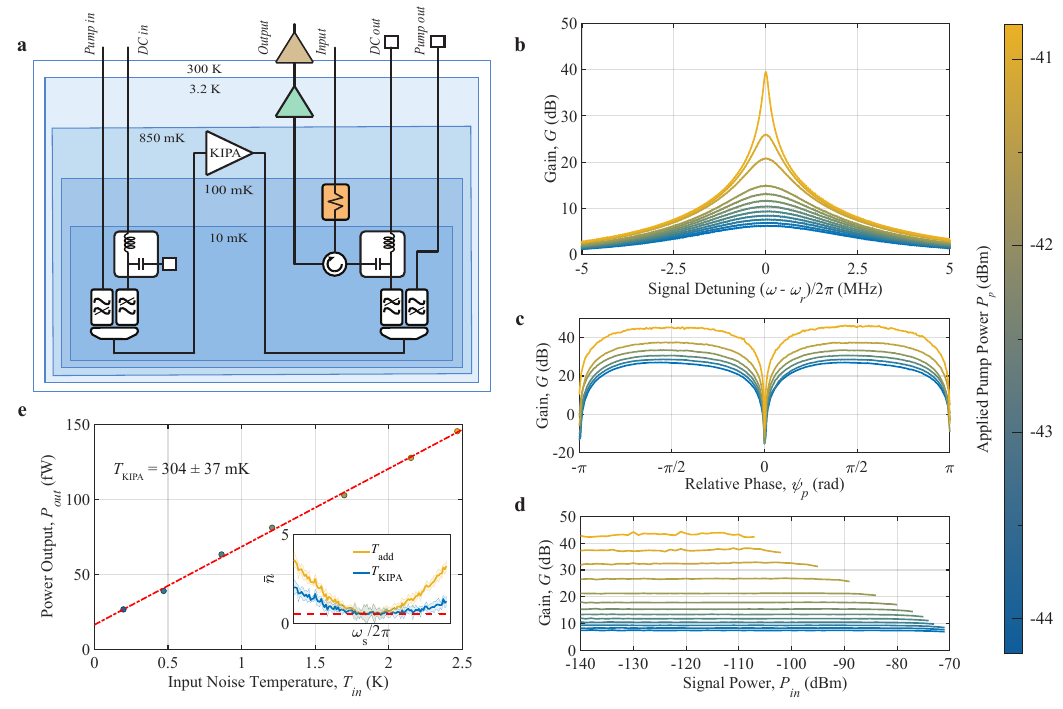}
	\caption{\textbf{KIPA operation at Still stage.} \textbf{(a) Simplified experimental setup}. Schematics of the experimental setup used to characterize the device in reflection at an operational temperature $T_\text{dev} = 850$~mK. The second port is intentionally mismatched at cryogenic temperature after the bias-tee. 
 \textbf{(b) Gain characteristics}. 
 Phase-insensitive gain of the KIPA measured in reflection as a function of signal frequency $\omega$ for various pump powers (see colorscale). Notably, the extracted gain-bandwidth product ($GBP$) achieved when measuring the device in reflection is $\sim 9$~MHz, which differs from the results in Fig.\ref{fig:characteristics}.
 \textbf{(c) Degenerate mode}. Phase-sensitive gain behavior obtained by operating at $\omega_s = \omega_p / 2$ and modulating the pump phase $\psi_p$. The maximum deamplification achieved is approximately $-15$~dB, in agreement with what reported in Appendix~\ref{app:maxsqueezing}. 
 \textbf{(d) Saturation power}. Phase-insensitive saturation power measured as signal power-dependent gain for different pump powers. The extracted saturation is $P_\text{1dB} = -63$~dBm, in line with the measurement performed at base temperature operation. 
 \textbf{(e) Noise performance}. Power output in correspondence of different VTS temperatures. The red dashed line represents the linear fit according to Eq.~\eqref{eq:pout}. The amplifier exhibits marginally worse noise performance when operated at 850~mK, with a noise temperature of $T_\text{KIPA} = 304 \pm 37$~mK, still preserving near-quantum-limited capabilities.}
	\label{fig:still}
\end{figure*}

\subsection{\label{app:themt}Noise temperature measurements: $T_\text{HEMT}$}

To accurately assess the contribution of the noise from the KIPA, it is essential to quantify the additional noise introduced by the HEMT low-noise temperature amplifier, denoted as $T_\text{HEMT}$. To accomplish this, we incorporate a 50~$\Omega$ through in the FSE to characterize the HEMT amplifier (LNF-LNC4\_8C, Low Noise Factory). The VTS temperature is systematically varied from 100~mK to 3~K in four steps, following the protocol outlined in \cite{Simbierowicz_2021}. We monitor the output power $P_\text{out}$ using a spectrum analyzer operating at a bandwidth of $B = 100$~Hz and averaging 20 waveforms to mitigate measurement noise. From the relation:
\begin{equation}
    \frac{P_\text{out}}{G_{\text{tot}} k_B B} = T_\text{in}^{(H)} + T_{\text{HEMT}} + \frac{T_{\text{bkg}}}{G_{\text{HEMT}}}. \label{eq:poutHEMT}
\end{equation}
Here, $T_\text{in}^{(H)}$ represents the injected noise from the VTS reaching the input HEMT and is given by:
\begin{equation}
    T_\text{in}^{(H)} = 2\eta_e \frac{\hbar \omega_s}{2 k_B} \coth{\frac{\hbar \omega_s}{2 k_B T_{\text{VTS}}}}.
\end{equation}
$T_{\text{HEMT}}$ accounts for the excess noise from the HEMT, while $T_{\text{bkg}}$ represents the room temperature background noise. Note that $T_\text{in}^{(H)}$ considers cryogenic electronic losses $\eta_e$ incurred twice as they appear in Fig.~\ref{fig:noise}(a). From the measurements, we extract the amplifier noise temperature $T_{\text{HEMT}}$ as 1.95~K, very close to the value provided by the manufacturer.

\subsection{\label{app:ttemp}Cascaded Transmission Efficiency in Fig.~\ref{fig:temperature}}

As reported in the main text, to characterize the KIPA at higher temperature, we perform a temperature sweep of the mixing chamber, as described in the main text. When warming up the system, we need to consider the contribution of noise introduced by cryogenic electronic devices operating at higher temperatures. In fact, for each loss channel of the $i$-th cryogenic electronic component, the noise propagation equation can be expressed as:
\begin{equation}
    N_{i} = \eta_i N_{i-1} + (1-\eta_i) N_{\text{e}_i}. \label{eq:Nei}
\end{equation}
Here, $N_{\text{e}_i}(T_{\text{e}_i})~=~1/[\exp{(\hbar \omega_s / k_B T_{\text{e}_i})}-1]$. In case of electronics operating at $T_{\text{e}_i} = 10$~mK, this simplifies to $N_{i} \approx \eta_i N_{i-1}$. However, when the electronics temperature $T_{\text{e}_i}$ exceeds few hundreds of millikelvin, the second term in Eq.~\eqref{eq:Nei} cannot be neglected, in particular for relatively high insertion loss devices.

Taking into account the losses originating from the electronics, the insertion loss of the amplifier, and all attenuators influencing the noise estimation, the term representing the added noise temperature in Eq.~\eqref{eq:pout} transforms into the expression detailed in Eq.~\eqref{eq:tadd}. From this formulation, we derive the noise temperature of the KIPA ($T_\text{KIPA}$).

\subsection{\label{app:still}Operation at Still stage}

To confirm that the KIPA maintains quantum noise limited performance when operated at higher temperatures, we conducted a characterization of the KIPA anchoring it on the ``still'' plate of a dilution cryostat with an effective temperature $T_\text{dev} = T_\text{Still} = 850$~mK. From an operational perspective, the still stage offers a much larger cooling power than the mixing chamber, all while keeping temperatures low enough to prevent the introduction of a relevant quasi-particle density in the NbN KIPA.


In this measurement, as illustrated in Fig.~\ref{fig:still}(a), we install the KIPA in a cryogenic system equipped with a VTS to investigate the noise performance of our device while operating it in a reflection configuration. We achieve this by introducing a mismatch on the bias tee with a 0~$\Omega$ termination.  As shown in Fig.~\ref{fig:still}(b-d), key parameters such as gain, phase-sensitivity, and saturation power are minimally affected by the higher operation temperature. The KIPA demonstrates a gain exceeding 20~dB, clear phase-sensitivity in the degenerate mode, and a saturation power of $P_\text{1dB} = -63$~dBm. However, a reduction in the gain bandwidth-product is observed, shifting from 20~MHz (in transmission) to 9~MHz (in reflection), likely associated with an imperfect mismatch of the second port.

Concerning noise performance, minimal differences are noted when compared to operation at base temperature, consistent with the findings in Ref.~\cite{Xu_2023b}. The estimated noise, obtained using the same procedure outlined in the main text, is $T_\text{KIPA} = 304 \pm 37$~mK, equivalent to $\bar{n} = 0.60 \pm 0.13$ added photons [see Fig.~\ref{fig:still}(e)]. Despite being slightly higher than the value obtained at base temperature, the device exhibits remarkable robustness to environmental temperature fluctuations when operated at 850~mK, attributable to its large internal quality factor.

\bibliography{Bibliography_General.bib,Bibliography_Cryogenics.bib,Bibliography_SNSPD.bib,Bibliography_SuperconductingCircuits.bib,Bibliography_SCresonators.bib,Bibliography_PA.bib,Bibliography_JPA.bib,Bibliography_KIPA.bib,Bibliography_TWPA.bib,Bibliography_QuantumDots.bib,Bibliography_Qubits.bib}

\end{document}